\documentclass[aps, amsmath, amssymb, 10pt, pra, twocolumn, longbibliography, superscriptaddress]{revtex4-1}

\usepackage{algcompatible}
\usepackage{amsmath}
\usepackage{amsthm}
\usepackage{bm}
\usepackage{braket}
\usepackage{subfigure}
\usepackage[dvipdfmx]{graphicx}
\usepackage{float}
\usepackage{url}
\usepackage{xcolor}
\usepackage{multirow}
\usepackage[colorlinks=true,urlcolor=blue,citecolor=blue,linkcolor=blue]{hyperref}
\usepackage{cleveref}
\usepackage{here}
\usepackage{algorithm}
\usepackage{physics}
\usepackage[noend]{algpseudocode}
\usepackage{comment}

\crefname{figure}{Fig.}{Figs.}
\crefname{equation}{Eq.}{Eqs.}
\crefname{section}{Sec.}{Sec.}
\Crefname{figure}{Figure}{Figures}
\Crefname{equation}{Equation}{Equations}
\Crefname{section}{Section}{Sections}

\makeatletter
\newcommand{\subalign}[1]{%
  \vcenter{%
    \Let@ \restore@math@cr \default@tag
    \baselineskip\fontdimen10 \scriptfont\tw@
    \advance\baselineskip\fontdimen12 \scriptfont\tw@
    \lineskip\thr@@\fontdimen8 \scriptfont\thr@@
    \lineskiplimit\lineskip
    \ialign{\hfil$\m@th\scriptstyle##$&$\m@th\scriptstyle{}##$\hfil\crcr
      #1\crcr
    }%
  }%
}
\makeatother

\begin{document}

\title{Floquet Analysis of Frequency Collisions}

\author{Kentaro Heya}
\affiliation{IBM Quantum, IBM T.J. Watson Research Center, Yorktown Heights, New York 10598, USA}
\email{Kentaro.Heya1@ibm.com}


\author{Moein Malekakhlagh}
\affiliation{IBM Quantum, IBM T.J. Watson Research Center, Yorktown Heights, New York 10598, USA}

\author{Seth Merkel}
\affiliation{IBM Quantum, IBM T.J. Watson Research Center, Yorktown Heights, New York 10598, USA}

\author{Naoki Kanazawa}
\affiliation{IBM  Quantum, IBM Research Tokyo,  19-21  Nihonbashi  Hakozaki-cho,  Chuo-ku,  Tokyo,  103-8510,  Japan}

\author{Emily Pritchett}
\affiliation{IBM Quantum, IBM T.J. Watson Research Center, Yorktown Heights, New York 10598, USA}

\begin{abstract}
Implementation of high-fidelity gate operations on integrated-qubit systems is of vital importance for fault-tolerant quantum computation.
Qubit frequency allocation is an essential part of improving control fidelity.
A metric for qubit frequency allocation, frequency collision, has been proposed on simple systems of only a few qubits driven by a mono-modal microwave drive.
However, frequency allocation for quantum processors for more advanced purposes, such as quantum error correction, needs further investigation.
In this study, we propose a Floquet analysis of frequency collisions.
The key to our proposed method is a reinterpretation of frequency collisions as an unintended degeneracy of Floquet states, which allows a collision analysis on more complex systems with many qubits driven by multi-modal microwave drives.
Although the Floquet state is defined in an infinite-dimensional Hilbert space, we develop algorithms, based on operation perturbation theory, to truncate the Hilbert space down to the optimal computational complexity.
In particular, we show that the computational complexity of the collision analysis for a sparse qubit lattice is linear with the number of qubits.
Finally, we demonstrate our proposed method on Cross-Resonance based experimental protocols.
We first study the Cross-Resonance gate in an isolated three-qubit system, where the effectiveness of our method is verified by comparing it with previous studies.
We next consider the more complex problem of syndrome extraction in the heavy-hexagon code~\cite{PhysRevX.10.011022}.
Our proposed method advances our understanding of quantum control for quantum processors and contributes to their improved design and control.
\end{abstract}

\maketitle

\section{Introduction}
There has been remarkable progress in quantum computer engineering toward the realization of large-scale fault-tolerant quantum computation~\cite{brayvi2022}.
In particular, superconducting circuits are among the most promising quantum platforms due to their compatibility with conventional transistor fabrication processes,  microwave control technology in communication wavelength bands~\cite{nakamura1999coherent, wallraff2004strong, majer2007coupling}, and progress towards long-range 3D integration~\cite{Foxen2018, Gold2021, Kjaergaard2020}.
Superconducting quantum processors with several hundred qubits~\cite{nation2022suppressing} allow demonstrations of quantum error correction~\cite{corcoles2015demonstration, PhysRevLett.119.180501, andersen2020repeated, andersen2020repeated, krinner2021realizing, ai2021exponential, PhysRevLett.128.110504, marques2022logical, PhysRevLett.129.030501, neeraja2022matching}.
However, the number of physical qubits required for fault-tolerant quantum computing is strongly dependent on physical gate fidelity.
It is therefore imperative to further improve physical gate fidelity.

Superconducting circuits can be broadly classified into those with and without a tunable Josephson junction~\cite{PhysRevApplied.8.044003}.
Tunable-circuit architectures allow dynamic control of circuit parameters and are generally superior in terms of execution time of entangling gates and system flexibility, but they suffer from sensitivity to external flux noise~\cite{dicarlo2009demonstration, PhysRevLett.111.080502, PhysRevApplied.6.064007, PhysRevApplied.10.034050, PhysRevLett.125.240503, PhysRevX.11.021058, PhysRevLett.127.080505}.
The fixed-circuit architectures, on the other hand, use fixed circuit parameters, resulting in a relatively simple implementation with excellent scalability, stability, and coherence~\cite{chow2013microwave, PhysRevLett.109.060501, PhysRevApplied.14.044039, PRXQuantum.2.040336, PhysRevLett.127.200502, PhysRevLett.129.060501, PhysRevApplied.16.054041, PRXQuantum.3.040322, nguyen2022programmable, shirai2023all}.
However, fixed-circuit architectures lack flexibility as their performance is highly dependent on designed circuit parameter values.
Qubit frequency allocation is therefore of great importance, especially for fixed-frequency architectures.

Frequency collisions~\cite{Malekakhlagh_2020, Brink2018DeviceCF, PRXQuantum.3.020301, doi:10.1126/sciadv.abi6690, kim2022effects} were introduced as a guideline for designing circuits that use the cross-resonance~(CR) gate~\cite{rigetti2010fully, Sheldon_2016, Magesan_2020}, a widely-used microwave-activated entangling gate amenable to fixed-frequency architectures.
Frequency collisions can be estimated numerically by time-domain simulation of the full CR Hamiltonian ~\cite{doi:10.1126/sciadv.abi6690} or analytically using time-independent or time-dependent Schrieffer-Wolff perturbation under the Rotating-Wave Approximation~(RWA) of the drive Hamiltonian~\cite{PhysRev.149.491, Magesan_2020, Tripathi_2019, malekakhlagh2021mitigating}.
Ref.~\cite{Malekakhlagh_2020} characterizes frequency collisions for the standard CR gate with consideration of a spectator qubit by looking into the poles of the effective Hamiltonian while sweeping system parameters.
Effective and adequate for isolated systems, numerical approaches are impractical for larger systems, and  time-independent perturbation under RWA is only valid for time-periodic mono-modal Hamiltonian.
In general, for a multi-modal Hamiltonian, there is no obvious choice of a rotating frame that makes the Hamiltonian time-independent~\cite{Wei2023}.
Therefore, the standard methods are not applicable, for example, to simultaneous CR gates~\cite{neeraja2022matching}, other multi-modal gates such as cross-cross resonance gate~\cite{PRXQuantum.2.040336}, or systems where interactions with detuned spectator qubits are important~\cite{Wei2023}.
In Ref.~\cite{PRXQuantum.3.020301}, on the other hand, an analysis of frequency collisions in flux-activated entangling gates~\cite{PhysRevApplied.6.064007} reveals that collisions are caused by accidental degeneracies of energy levels during flux-bias sweeps, both in isolation and in parallel with simultaneous flux-bias sweeps.
Note, however, this analysis applies only when the Hamiltonian is approximately time-independent during the gate.

In this paper, we redefine frequency collision by means of a more general and unified framework based on Floquet theory~\cite{floquet1883equations}.
Floquet theory introduces frequency-domain analysis to frequency collisions, which are commonly analysed in the time domain~\cite{hertzberg2020laserannealing, doi:10.1126/sciadv.abi6690}.
It characterizes frequency collisions as the breakdown of strong-dispersive condition of the Floquet Hamiltonian in the desired operation basis.
To distinguish such frequency-domain frequency collisions from the previous notions, we call them ``Floquet collisions''.
The degree of Floquet collision is quantified as a collision angle between Floquet states, and the degree of fidelity deterioration caused by Floquet collisions can also be estimated from the collision angle and the effective coupling strength between the colliding Floquet states.
Because the Floquet Hamiltonian has infinite dimension, we propose a perturbative approach~\cite{primas1963generalized} to derive the collision angles with a finite computational cost.
We also discuss a relationship between Floquet collision and qubit lattice structure in real space, and propose an efficient collision analysis method for the sparse qubit lattice that requires only a linear computational cost with respect to the number of qubits.

We demonstrate our Floquet-based collision analysis on analytical and numerical simulations.
First, for CR gates on an isolated-transmon system,
we provide analytical solutions of Floquet collisions in terms of collision angles between Floquet states and re-examine the frequency collisions in Ref.~\cite{malekakhlagh2021mitigating}.
Our Floquet-based collision analysis can be applied to an arbitrary parameter region, and allows for the calculation of the intensity and bounds of collisions, which extends the discussion in Refs.~\cite{hertzberg2020laserannealing, doi:10.1126/sciadv.abi6690}.
Second, we calculate and visualize the collision angles numerically while sweeping various system parameters such as qubit frequencies, CR drive amplitude, and rotary tone drive amplitude~\cite{Sheldon_2016, PRXQuantum.1.020318}.
Our numerical simulations reveals novel collisions originating from microwave drive, which have not been reported before.
These findings will contribute to the optimal pulse shaping of the CR gate in the future.
Third, we consider a more complex system of error correction code.
Here, we focus on syndrome extraction in heavy-hexagon codes~\cite{PhysRevX.10.011022, neeraja2022matching}, quantitatively estimating the difficulty of system frequency allocation.
Our measures provide an approximate estimate of the computational cost to improve the system frequency allocation for large-scale error correction in the future.

The rest of the main text is organized as follows:
In \cref{sec:floquet}, we review Floquet theory, and our reinterpretation of frequency collisions.
In \cref{sec:perturbation}, we review generalized perturbation theory and apply it to find Floquet collisions.
We also discuss the relationship between Floquet collision and distance in Floquet and real space.
In \cref{sec:complexity}, we summarize our Floquet-based collision analysis, and estimate the computational complexity of collision analysis.
In \cref{sec:demonstration}, we demonstrate our Floquet-based collision analysis by applying it to actual experimental conditions.
Lastly, in \cref{sec:conclusion}, we summarize our results and provide directions for future work.

\section{Frequency Collision and Floquet theory} \label{sec:floquet}
Floquet theory~\cite{floquet1883equations} applies to linear differential equations with time-periodic generators, having a wide range of applications such as stability analysis~\cite{dugundji1983some, peters1994fast}, chemistry~\cite{PhysRevE.79.051131}, and material physics~\cite{oka2019floquet, de2019floquet}.
In the context of quantum mechanics, a time-periodic Hamiltonian with Continuous Wave~(CW) drive is a prime example of a time-periodic generator~\cite{PhysRev.138.B979, levante1995formalized}.
We can also find recent applications to quantum information science such as analyzing time crystals~\cite{PhysRevLett.117.090402}, gate calibration~\cite{arute2020observation}, controlling driven qubits~\cite{PhysRevApplied.17.064006, PRXQuantum.3.040322, nguyen2022programmable},
protecting qubits from noise~\cite{PhysRevApplied.14.054033, PhysRevApplied.15.034065, PRXQuantum.3.020337}, and 
improving a quantum parametric amplifier~\cite{PRXQuantum.3.020306}.

In this section, we apply Floquet theory to the analysis of frequency collisions.
Using Floquet theory, we can transform a finite-dimensional periodic time-dependent Hamiltonian into an infinite-dimensional time-independent Floquet Hamiltonian, instead of the common practice of transforming into a rotating frame under the RWA.
Floquet theory enables analyzing frequency collisions in a multi-modal Hamiltonian~\cite{PhysRevA.102.062408, PRXQuantum.2.040336, PhysRevApplied.17.064006, PhysRevApplied.15.034065, PhysRevApplied.14.054033, PRXQuantum.3.020337, PRXQuantum.3.040322} and provides deeper insights into frequency collisions.

\subsection{Floquet theory} \label{sec:floquet_theory}
\begin{figure*}
    \centering
	\includegraphics[width=0.9\textwidth]{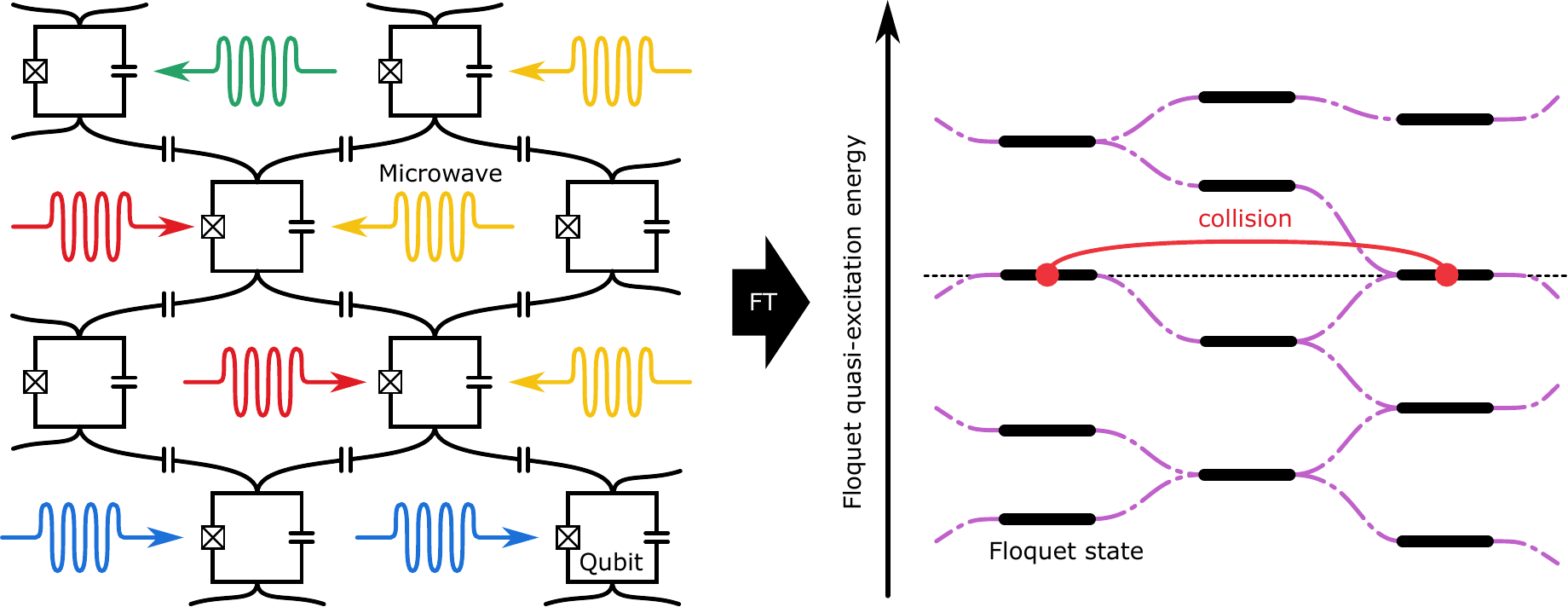}
    \caption{
    Concept of our Floquet-based collision analysis.
    The left figure shows a general quantum control schematic.
    The qubits align in a planar lattice, and multiple microwaves are simultaneously irradiated.
    Some of the microwaves can be resonant with each other.
    Using Floquet theory and applying the Fourier transform~(FT) to the Schrodinger equation corresponding to the left figure, we obtain a Floquet Hamiltonian and a corresponding energy level diagram shown on the right.
    The black bold and purple dotted lines represent Floquet states and their couplings, respectively.
    If the Floquet Hamiltonian satisfies the strongly dispersive condition in the operation basis described in \cref{sec:floquet_collision}, we can control our system sufficiently.
    On the other hand, if Floquet states are degenerate, as shown in the right figure, Floquet collision occurs and the control fidelity deteriorates significantly.
    }
    \label{fig:concept}
\end{figure*}

Let us assume a $d$-dimensional system Hamiltonian $H(t)$.
The Schr\"{o}dinger equation in natural units is given as:
\begin{align}
\qty(H(t) - i\dv{t}) \Phi(t) = 0 \;,
\label{eq:schrodinger_eq}
\end{align}
where $\Phi(t)$ represents the fundamental solution matrix.
According to Floquet theory, the fundamental solution matrix of a differential equation reads:
\begin{align}
\Phi(t) &= e^{-iEt}\Psi(t) \;,
\label{eq:fundamental_solusion_matrix}
\end{align}
where $\Psi(t)$ is a $d$-dimensional square matrix and $E$ is a $d$-dimensional diagonal matrix.
$\Psi(t)$ has the same time periodicity as $H(t)$ and can be decomposed by the discrete Fourier series expansion as:
\begin{align}
H(t) &=\sum_{\va{n}\in\mathbb{Z}^{\norm{\va{\omega}}_0}} H^{\qty(\va{n})} e^{i\va{n}\vdot\va*{\omega}t} \;,  
\label{eq:fourier_series_H_expansion} \\
\Psi(t) &=\sum_{\va{n}\in\mathbb{Z}^{\norm{\va{\omega}}_0}} \Psi^{\qty(\va{n})} e^{i\va{n}\vdot\va*{\omega}t}\;,
\label{eq:fourier_series_Psi_expansion}
\end{align}
where $A^{\qty(\va{n})}$ represents the $\va{n}$th coefficient of the discrete Fourier series expansion of the matrix $A$.
Then, we obtain the following eigenequation from the Schr\"{o}dinger equation:
\begin{align}
E_{ii} \Psi^{\qty(\va{n})}_{ij} = \sum_{k,\va{m}}
\qty{
    H^{\qty(\va{n}-\va{m})}_{ik} + 
    \qty(\va{n}\vdot\va*{\omega})
    \delta_{ik}\delta_{\va{n}\va{m}}
}
\Psi^{\qty(\va{m})}_{kj}\;,
\label{eq:floquet_eigen_equation}
\end{align}
where $A_{ij}$ represents the $(i,j)$-elements of the matrix $A$.
Note that \cref{eq:floquet_eigen_equation} is invariant except for a constant offset for any translation operation $\va{r}$ on the vector $\va{n}$.

\Cref{eq:floquet_eigen_equation}  can be interpreted as a time-independent energy level diagram in infinite dimension.
In the following, we refer to this extended energy level diagram as the Floquet Hamiltonian and the subspace associated with index $\va{n}$ as the Brillouin zone $\mathrm{BZ}\qty(\va{n})$~\cite{PhysRevA.7.2203}.
The state $\ket{\psi}$ in $\mathrm{BZ}\qty(\va{n})$ is denoted as $\ket{\psi;\va{n}}$.
We refer to the eigenvalues and eigenvectors of the Floquet Hamiltonian as Floquet quasi-excitation energies and Floquet states, respectively.
The system time evolution $U(t';t)$ can be reconstructed as:
\begin{align}
U(t';t)
&=\Phi(t')\Phi(t)^\dagger \\
&=e^{-iEt'}\Psi(t')\Psi^\dagger(t)e^{iEt} \;.
\label{eq:gate_fidelity}
\end{align}

\subsection{Frequency collision} \label{sec:floquet_collision}
\begin{figure}
    \centering
	\includegraphics[width=0.375\textwidth]{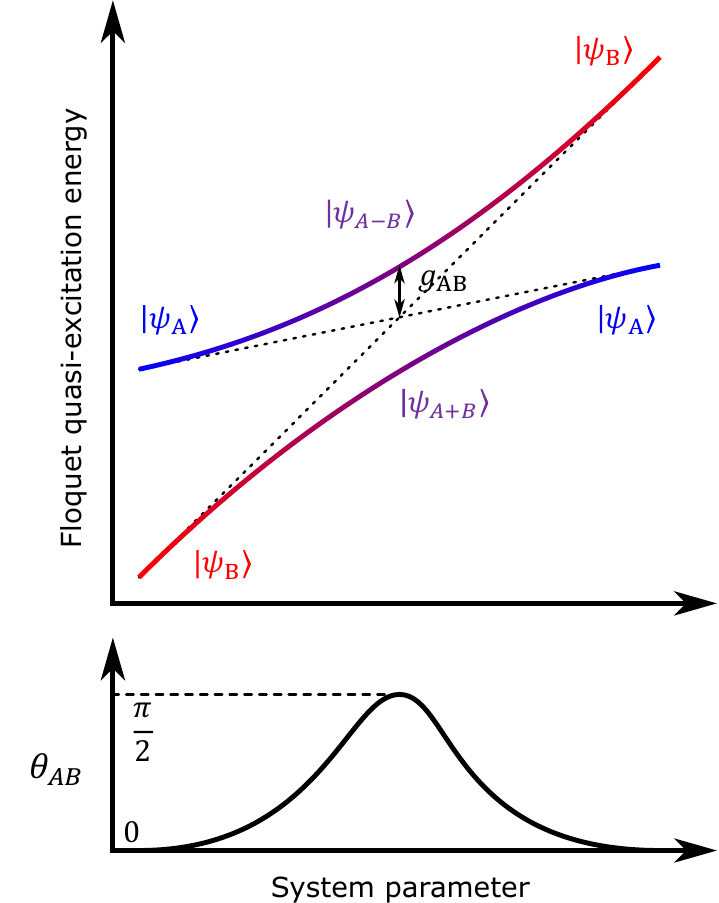}
    \caption{
    (Top) A schematic for anti-crossing between Floquet states $\ket{\psi_{A,B}}$.
    The top figure shows how Floquet quasi-excitation energy changes while sweeping a system parameter.
    We can find that an anti-crossing occurs near a degenerate point of two Floquet states, where Floquet the quasi-excitation energy acquires an energy gap proportional to an effective coupling strength between the Floquet states.
    (Bottom) Dependence of collision angle $\theta_{\mathrm{AB}}$ between the Floquet states $\ket{\psi_{\mathrm{A},\mathrm{B}}}$ while sweeping system parameter.
    The collision angle becomes $\pi/2$ at the degenerate point and close to $0$ under a strong-dispersive condition.
    }
    \label{fig:anti-cross}
\end{figure}

In the context of quantum gates, a frequency collision refers to proximity to unwanted resonances that are detrimental to gate performance.
The frequency collision can be static or dynamic, depending on the nature of the underlying transition and the role of drive photons.
In a perturbative analysis, the position of frequency collisions can be identified as poles, while the collision width and nature is determined by the makeup of transition matrix elements in the numerator~\cite{Malekakhlagh_2020, Brink2018DeviceCF, hertzberg2020laserannealing, kim2022effects, doi:10.1126/sciadv.abi6690, PRXQuantum.3.020301}.
Using Floquet theory, collisions can be formulated more rigorously as a breakdown of the strong-dispersive condition~\cite{PhysRevA.74.042318, schuster2007resolving} in the Floquet Hamiltonian. 
In this section, we calculate a fidelity between ideal propagators and those in the presence of a collision.
This gives us a metric to numerically evaluate the impact of collisions on gate performance.

To describe a given quantum gate, we pick a suitable Floquet basis $\qty{\psi_i}$, which we generally refer to as the operation basis.
In the context of the CR gate, this is often called the block-diagonal basis.
To illustrate our method, consider the simple case of a Floquet subspace Hamiltonian spanned by two Floquet states $\ket{\psi_{\mathrm{A}};n+m}$, $\ket{\psi_{\mathrm{B}};n}$ in the operation basis with a drive frequency $\omega_d$ as follows:
\begin{align}
H_{\mathrm{AB}}=\Delta_{\mathrm{AB}} \frac{Z_{\mathrm{AB}}}{2} + 2g_{\mathrm{AB}} \frac{X_{\mathrm{AB}}}{2}\;,
\end{align}
where $\Delta_{\mathrm{AB}}$ and $g_{\mathrm{AB}}$ represent the energy detuning and coupling strength between the Floquet states, and $X_{\mathrm{AB}}$ and $Z_{\mathrm{AB}}$ represents the Pauli $X$ and $Z$ operator defined on the subspace.
The first and second terms represent the diagonal and non-diagonal elements of the Floquet Hamiltonian in the operation basis, respectively.

The matrix can be exactly diagonalized by the following rotational matrix:
\begin{align}
R_Y(\theta)&\equiv\exp\qty(-i\theta\frac{Y_{\mathrm{AB}}}{2})\;, \\
\theta&=\arctan{\qty(\qty|\frac{2g_{\mathrm{AB}}}{\Delta_{\mathrm{AB}}}|)}\;,
\end{align}
where $Y_{\mathrm{AB}}$ and $\theta$ represent the Pauli $Y$ operator and the collision angle between the Floquet states, respectively.
Thus, the diagonalized Floquet subspace Hamiltonian $E_{\mathrm{AB}}$ is found as:
\begin{align}
E_{\mathrm{AB}}&=R_Y(\theta)H_{\mathrm{F}}R_Y(-\theta)=r\frac{Z_{\mathrm{AB}}}{2}\;,
\label{eq:floquet_quasi_excitation_energy_simple}
\end{align}
where $r=\sqrt{\Delta_{\mathrm{AB}}^2+4g_{\mathrm{AB}}^2}$ represents the energy detuning between the Floquet states.
From \cref{eq:fourier_series_Psi_expansion}, the periodic functions on the subspace spanned by the Floquet states are reconstructed as:
\begin{align}
\Psi(t)
=R_Y (\theta) R_Z(m \omega_d t)
\mqty(\ket{\psi_{\mathrm{A}}} \\ \ket{\psi_{\mathrm{B}}})\;,
\end{align}
where a global phase is omitted for simplicity.
Then, from \cref{eq:fundamental_solusion_matrix} and \cref{eq:gate_fidelity}, the propagator from time $0$ to $T$ is found as:
\begin{align}
U_{\mathrm{AB}}(T;0) 
&= R_Z(rT)\Psi(T)\Psi^\dagger(0) \\
&= R_Z(rT) R_Y(\theta) R_Z(m \omega_d T) R_Y(-\theta) \;.
\label{eq:propagator_simple}
\end{align}
Without impact from a collision, we would expect $U_{\mathrm{AB}}(T;0) 
=  R_Z(\Delta_{\mathrm{AB}} T) R_Z(m \omega_d T)$, so from \cref{eq:propagator_simple}, Floquet collisions cause two types of gate errors simultaneously: axial rotation $R_Y(\theta)$ and phase shift $R_Z((r-\Delta_{\mathrm{AB}})T)$.

In the ideal case, Floquet Hamiltonian satisfies the strong-dispersive condition $g_{\mathrm{AB}}\ll\Delta_{\mathrm{AB}}$ or $\theta_{\mathrm{AB}}\sim0$ in the operation basis.
The ideal subspace propagator $V_{\mathrm{AB}}(T;0)$ designed under the strong-dispersive condition is obtained by linearly approximating the real subspace propagator $U_{\mathrm{AB}}(T;0)$ as:
\begin{align}
V_{\mathrm{AB}}(T;0)
&=R_Z\qty(v_{0}) \mqty(v_{1} & v_{3} \\ v_{3} & v_{2}) \;,
\end{align}
where $v_{i}$ read
\begin{align}
v_{0} &= \Delta_{\mathrm{AB}}T \qty{1+\frac{1}{2}\qty(\frac{2g_{\mathrm{AB}}}{\Delta_{\mathrm{AB}}})^2} \;, \\
v_{1} &= 1 + e^{im\omega_d T} \qty(\frac{2g_{\mathrm{AB}}}{\Delta_{\mathrm{AB}}})^2 \;, \\
v_{2} &= e^{im\omega_d T} + \qty(\frac{2g_{\mathrm{AB}}}{\Delta_{\mathrm{AB}}})^2 \;, \\
v_{3} &= \qty(1-e^{im\omega_d T})\qty(\frac{2g_{\mathrm{AB}}}{\Delta_{\mathrm{AB}}}) \;.
\end{align}
Thus, the inner product of the gates in the subspace is written as follows:
\begin{align}
f_{\mathrm{AB}}
&= \frac{1}{2}\mathrm{Tr}\qty[U_{\mathrm{AB}}(0;T) V_{\mathrm{AB}}^\dagger(0;T)] \\
&\geq \frac{1}{2}\mathrm{Tr}\qty[U_{\mathrm{AB}}(0;T) R_Z\qty(-(\Delta_{\mathrm{AB}} + m\omega_d)T)] \\
&=
\cos^2\qty(\frac{\theta}{2})
\cos\qty(\frac{\delta r T}{2}) \nonumber \\
&~~ + \sin^2\qty(\frac{\theta}{2})
\cos\qty(\qty(\frac{\delta r}{2}+m\omega_d)T)\;,
\label{eq:sub_fidelity}
\end{align}
where $\delta r=r - \Delta_{\mathrm{AB}}$.

Let us assume a $D$-dimensional Hilbert space, where we have no Floquet collision except for the two Floquet states $\ket{\psi_{\mathrm{A},\mathrm{B}}}$.
Then, the Floquet Hamiltonian of the whole space is expressed as follows:
\begin{align}
\mathcal{H} &= \qty(
\begin{array}{c|ccc} 
H_\perp & H_{\mathrm{int}} \\ \hline
H_{\mathrm{int}} & H_{\mathrm{AB}}
\end{array})
\;,
\end{align}
where $H_{\perp}$ and $H_{\mathrm{int}}$ represent the Floquet Hamiltonian in the outside of the subspace and the interaction term between the inside and the outside of the subspace, respectively.
From \cref{eq:floquet_quasi_excitation_energy_simple}, we can diagonalize the Floquet subspace Hamiltonian $H_{\mathrm{AB}}$ as:
\begin{align}
E
&=\qty(I_{\perp}\otimes R_Y(\theta))H\qty(I_{\perp}\otimes R_Y(-\theta)) \;, \\
&=\qty(
\begin{array}{c|ccc} 
H_\perp & H_{\mathrm{int}}R_Y(-\theta) \\ \hline
R_Y(\theta)H_{\mathrm{int}} & E_{\mathrm{AB}}
\end{array}) \;,
\end{align}
where $I_{\perp}$ represent the identity operator in the outside of the subspace.
Because the subspace-diagonalized Floquet Hamiltonian $E$ satisfies the strong-dispersive condition, we can solve the time evolusion driven by $E$ perturbatively.
Thus, the inner product of the gates in the whole space is approximated as follows:
\begin{align}
f=\frac{D-2}{D}+\frac{2}{D}f_{\mathrm{AB}}+\order{g_\mathrm{AB}H^2_{\mathrm{int}}} \;.
\end{align}
Consequently, the entangling fidelity~\cite{NIELSEN2002249} derived from the Floquet collision is calculated as follows:
\begin{align}
\mathcal{F}(U,V)
\equiv \qty|f|^2 \;.
\label{eq:whole_fidelity}
\end{align}

In the above, we dealt with the case where two Floquet states are degenerate with each other.
However, in realistic cases as described in \cref{sec:demo_cr}, it is possible for multiple Floquet states to collide simultaneously under certain system parameters.
In such a situation, we can follow the same procedure by diagonalizing the subspace spanned by the colliding Floquet states.
We also have a simpler approach to account for individual Floquet collisions independently, and estimate approximate lower bounds of the gate fidelity.
From a derivative of Gershgorin circle theorem~\cite{gershgorin1931uber}, we can bound the shift of the Floquet quasi-excitation energy detuning $\delta r_{ij}$, and the collision angles $\delta \theta_{ij}$ between $i$th and $j$th Floquet states, defined on the Floquet subspace $\mathcal{S}$ spanned by simultaneously colliding Floquet states, as follows:
\begin{align}
\abs{\delta r_{ij}} &\leq \delta r_{ij}^{\max} = \sum_{k \in \mathcal{S}} \qty(\abs{g_{ik}} + \abs{g_{jk}}) \;, \\
\abs{\theta_{ij}} &\leq \theta_{ij}^{\max}=\arctan{\qty(\qty|\frac{\delta r_{ij}}{\Delta_{ij}}|)} \;,
\label{eq:gershgorin}
\end{align}
where $g_{ij}$ and $\Delta_{ij}$ are the coupling strength and the energy detuning between the $i$th and $j$th Floquet states.
Note that we defined $g_{ii}=g_{jj}=0$.
From \cref{eq:sub_fidelity}, \cref{eq:whole_fidelity}, and \cref{eq:gershgorin}, we can estimate the lower bound of the control fidelity under such simultaneous Floquet collisions.

\section{Collision order and Generalized perturbation theory} \label{sec:perturbation}
To find Floquet collisions, we need to search for the degeneracies of the Floquet Hamiltonian in the operation basis.
Note that degeneracies not only occur between directly coupled Floquet states, but also between remote Floquet states via intermediate coupling paths between them.
We compute such degeneracies by diagonalizing the Floquet Hamiltonian.
If the components of two Floquet eigenstates are hybridized beyond a non-negligible fraction~(threshold), we identify it as a Floquet collision.
However, as shown in \cref{eq:fourier_series_H_expansion}, Floquet Hamiltonians are generally infinite dimensional, and exact diagonalization is impossible with finite computational cost.
However, depending on the complexity of the problem, we can perform such a diagonalization analytically using perturbation theory.

Perturbation theory gives a method for finding approximate order-by-order solutions that are slightly different from a solvable unperturbed system.
In quantum mechanics, various perturbative analyses of Hamiltonian dynamics have been proposed such as Rayleigh-Schr\"{o}dinger perturbation theory~\cite{rayleigh1896theory}, Schrieffer-Wolff perturbation
theory~\cite{PhysRev.149.491}, Dyson series~\cite{PhysRev.75.486}, Magnus expansion~\cite{magnus1954exponential}, average Hamiltonian theory~\cite{PhysRev.175.453}, and multi-scale perturbation theory~\cite{PhysRevLett.77.4114}.
We can also find several attempts of perturbative analysis on Floquet Hamiltonian in previous studies~\cite{guerin1999floquet, rodriguez2018floquet, PhysRevB.102.235114}.
In this section, we propose a method for evaluating Floquet collisions with finite computational cost by applying generalized perturbation theory~\cite{primas1963generalized} on the Floquet Hamiltonian.

\subsection{Generalized perturbation theory} \label{sec:perturbation_theory}
Generalized perturbation theory~\cite{primas1963generalized} is formulated in superoperator notation, and hence unifies the aforementioned perturbation theories.
We assume the following Hamiltonian:
\begin{align}
H=K+V \;,
\end{align}
where $K$ and $V$ correspond to the ``bare'' and ``perturbation'' terms, respectively, and generally do not commute with each other.
The goal of generalized perturbation theory is to transform the perturbation $V$ to commute with the bare term $K$ by finding an appropriate frame operator $G$ to satisfy the following equation:
\begin{align}
\comm{e^{\mathcal{G}}\qty(H)}{K} &= 0 \;,
\label{eq:frame_change}
\end{align}
where $\mathcal{G}(H)=\qty[G, H]$.

Let us assume the polynomial expansions of $G$ and $\mathcal{G}$ with respect to $V$ as follows:
\begin{align}
G&=\sum_{i=1}^{\infty} G_i~~(G_i\propto V^i)\;, \\
\mathcal{G}&=\sum_{i=1}^{\infty}\mathcal{G}_i~~\qty(\mathcal{G}_i(H)=\comm{G_i}{H})\;.
\end{align}
Then, the transformed Hamiltonian will be written as
\begin{align}
e^{\mathcal{G}}\qty(H)
&=\qty{\sum_{i=0}^{\infty}\frac{\qty(\sum_{j=1}^{\infty} \mathcal{G}_j)^i}{i!}}(H)\;,
\end{align}
and its polynomial expansions with respect to $V$ are written as follows:
\begin{align}
e^{\mathcal{G}}\qty(H)&=\sum_i H^{(i)}~~\qty(H^{(i)}\propto V^i)\;,
\end{align}
where $H^{(i)}$ represents a perturbatively expanded Hamiltonian and is written as follows:
\begin{align}
H^{(i=0)} &= K\;, \\
H^{(i=1)} &= \mathcal{G}_1 (K) + V \;, \\
H^{(i\geq2)} &=
\sum_{j=1}^{i}
\qty{
\frac{1}{j!}
\sum_{\subalign{\va{n}&\in\mathbb{N}^j\\\norm{\va{n}}_1&=i}}
\qty(\prod_{n\in\va{n}} \mathcal{G}_{n})(K)
} \nonumber \\
&+
\sum_{j=1}^{i-1}
\qty{
\frac{1}{j!}
\sum_{\subalign{\va{n}&\in\mathbb{N}^j\\\norm{\va{n}}_1&=i-1}}
\qty(\prod_{n\in\va{n}} \mathcal{G}_{n})(V)
}\;.
\label{eq:perturbative_floquet_ham}
\end{align}
We then enforce \cref{eq:frame_change} at each order:
\begin{align}
\comm{H^{(i)}}{K}=0\;,
\end{align}
from which the frame operators $G_i$ for $i\geq1$ are derived sequentially as follows:
\begin{align}
G_i&=\qty(\mathcal{D}_K\circ\mathcal{P}_K)\qty(H^{(i)} - \mathcal{G}_i(K))\;,
\label{eq:frame_operator}
\end{align}
where we use the spectrum decomposition
\begin{align}
K\equiv\sum_n \kappa_n K_n\;,
\end{align}
and the following superoperators
\begin{align}
\mathcal{D}_K(X)&\equiv\sum_{n\neq m} \frac{K_n X K_m}{\kappa_n - \kappa_m}\;, \\
\mathcal{P}_K(X)&\equiv X-\sum_n K_n X K_n \;.
\end{align}
From \cref{eq:perturbative_floquet_ham} and \cref{eq:frame_operator}, we note that the term $H^{(i)}-\mathcal{G}_i(K)$ consists of only $K$, $V$ and $\qty{G_j}_{j<i}$.

Suppose the frame operator $G^{(k)}$, truncated up to $k$th order in perturbation, is as follows:
\begin{align}
G^{(k)}&=\sum_{i=1}^{k} G_i\;.
\end{align}
The Hamiltonian transformed by $G^{(k)}$ is found as:
\begin{align}
e^{\mathcal{G}^{(k)}}(H)&=\sum_{i=0}^{k}H^{(i)} + \order{V^{k+1}}\;,
\label{eq:perturbative_diagonalization} \\
\mathcal{G}^{(k)}(H)&=\comm{G^{(k)}}{H}\;.
\end{align}
From \cref{eq:perturbative_diagonalization}, the frame change with operator $G^{(k)}$ can be regarded as a perturbative diagonalization up to the $k$th order.
Note that the difference between the diagonal terms, and the off-diagonal terms, of the perturbatively diagonalized Floquet Hamiltonian correspond to the energy detuning $\Delta^{(k)}_{ij}$, and the effective coupling strength $g^{(k)}_{ij}$, respectively:
\begin{align}
\Delta^{(k)}_{ij} &= e^{\mathcal{G}^{(k)}}(H)_{ii}-e^{\mathcal{G}^{(k)}}(H)_{jj}\;, \\
g^{(k)}_{ij} &= e^{\mathcal{G}^{(k)}}(H)_{ij}\;.
\end{align}
Therefore, we can verify the Floquet collisions between the $i$th and $j$th diagonal elements by a collision angle defined as:
\begin{align}
\theta^{(k)}_{ij} \equiv 
\arctan{\qty(\qty|\frac{2g^{(k)}_{ij}}{\Delta^{(k)}_{ij}}|)}\;.
\label{eq:floquet_quantization_angle}
\end{align}
In the following, when the $(k+1)$th-order effective coupling strength produced by $k$th-order perturbative diagonalization becomes non-negligible with respect to the energy detuning between them, we will call this a $(k+1)$th-order Floquet collision.

\subsection{Collision order and distance in Floquet space}\label{sec:floquet_space}
\begin{figure*}
    \centering
	\includegraphics[width=0.95\textwidth]{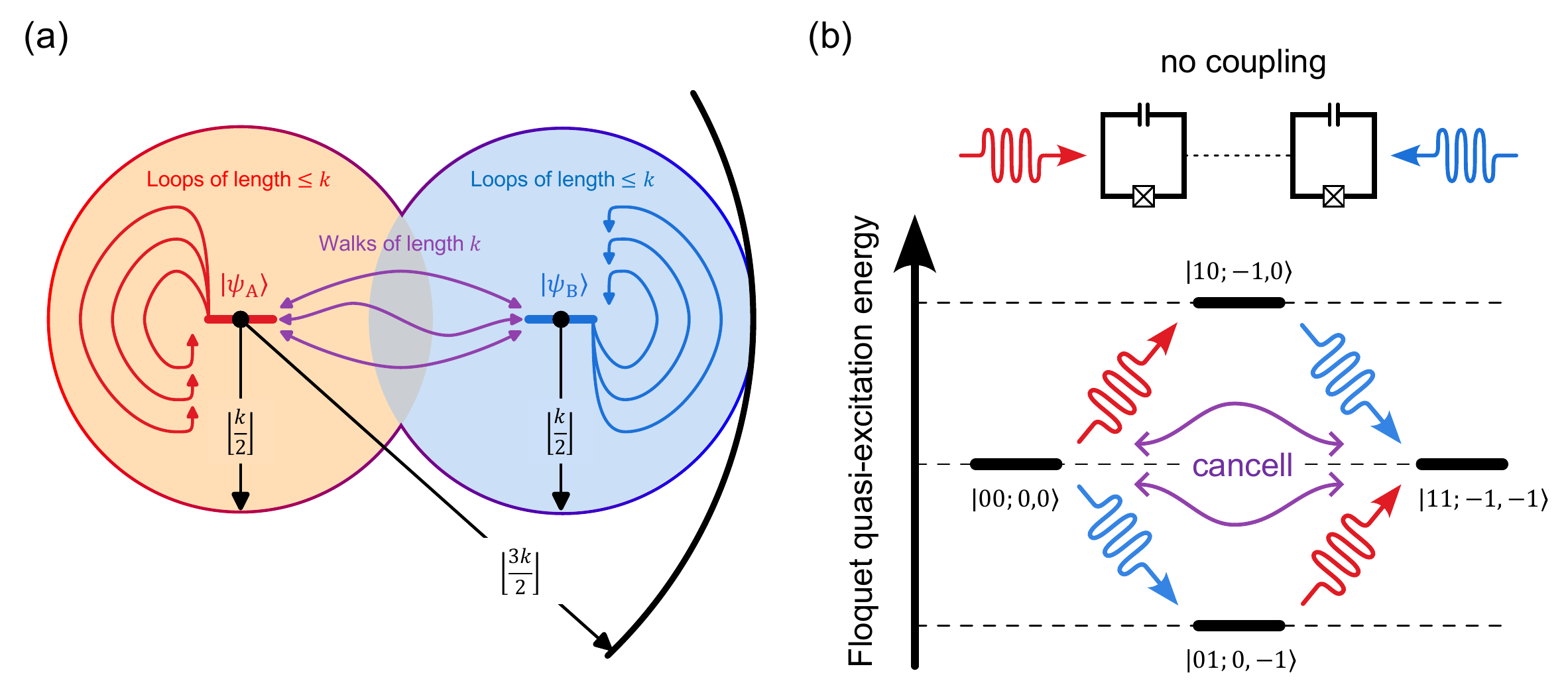}
    \caption{
    (a)
    Graphical interpretation of the $k$th-order perturbations acting on the Floquet states $\ket{\psi_{\mathrm{A},\mathrm{B}}}$.
    A loop of length $k$ around Floquet states $\ket{\psi_{\mathrm{A},\mathrm{B}}}$ and a walk of length $k$ between them on the Floquet Hamiltonian correspond to $k$th-order Floquet quasi-excitation energy shifts and effective coupling, respectively.
    To verify the $k$th-order Floquet collision, we compare the Floquet quasi-excitation energy detuning, accounting for up to $k$th-order energy shifts, to the $k$th-order effective coupling strength.
    Such loops and walks are contained in a region within distance $d\leq \lfloor{k/2\rfloor}$ of each state, such that we can use the Floquet subspace Hamiltonian corresponding to this region to verify the $k$th-order Floquet collision with finite computational cost.
    All Floquet collisions up to the $k$th-order caused on the Floquet state $\ket{\psi_{\mathrm{A}}}$ include only Floquet collisions with the Floquet states within distance $d\leq k$.
    Thus, all Floquet collisions caused on $\ket{\psi_{\mathrm{A}}}$ can be calculated with the Floquet subspace Hamiltonian corresponding to the region within distance $d\leq \lfloor{3k/2\rfloor}$ from $\ket{\psi_{\mathrm{A}}}$ in the Floquet Hamiltonian.
    (b)
    The circuit and the corresponding low-excitation Floquet energy level diagram of two isolated transmon qubits irradiated by independent microwave drives.
    Even though there is no qubit-qubit interaction, there is a walk of distance $2$ between states $\ket{0,0;0,0}$ and $\ket{1,1;-1,-1}$ mediated by $\ket{0,1;0,-1}$ and $\ket{1,0;-1,0}$ on the Floquet Hamiltonian.
    However, the effective couplings between them, mediated by two coupling walks, always cancel each other out.
    }
    \label{fig:graphical_perturbation}
\end{figure*}
We identified Floquet collisions as the breakdown of the strong-dispersive conditions in the Floquet Hamiltonian, which can be detected via the collision angles defined in \cref{eq:floquet_quantization_angle}.
However, the infinite dimensionality of the Floquet representation makes the exact diagonalization impossible.
Hence, a systematic truncation procedure is necessary.
In this subsection, we show that it is possible to verify the Floquet collisions below a certain order using only a finite dimensional Floquet subspace Hamiltonian.

Let us define a Floquet subspace $\mathcal{S}^{(k)}(\ket{\psi})$ spanned by Floquet states with non-zero effective couplings to a Floquet state $\ket{\psi}$ on the $k$th-order perturbatively diagonalized Floquet Hamiltonian $H^{(k)}$ as follows:
\begin{align}
\mathcal{S}^{(k)}(\ket{\psi})\equiv\mathrm{span}\qty(H^{(k)}\ket{\psi})\;,
\end{align}
where ``span'' represents the subspace spanned by all the Floquet states in the argument.
Here, we prove that the subspace $\mathcal{S}^{(k\leq)}(\ket{\psi})$ is encompassed in a $k$th-order Krylov subspace~\cite{hestenes1952methods} with the perturbation $V$ and the Floquet state $\ket{\psi}$ defined as follows:
\begin{align}
\mathbb{K}_k(V,\ket{\psi})\equiv\mathrm{span}\qty(\qty{V^j \ket{\psi}}_{j=0}^{k})\;.
\end{align}
From \cref{eq:perturbative_floquet_ham}, we can represent the $k$th-order perturbatively diagonalized Floquet Hamiltonian $H^{(k)}$ as follows:
\begin{align}
H^{(k)}=\mathrm{Poly}\qty(V^k, K^x, \mathcal{D}_K^y, \mathcal{P}_K^y)\;,
\end{align}
where $x,y\in\mathbb{N}$.
Since $K$ and $\mathcal{D}_K$ are diagonal in the operation basis, they do not change the dimension of $\mathcal{S}^{(k)}(\ket{\psi})$ such that:
\begin{align}
\mathcal{S}^{(k)}(\ket{\psi})=\mathrm{span}\qty(\mathrm{Poly}\qty(V^k, \mathcal{P}_K^y)\ket{\psi})\;.
\label{eq:span_space}
\end{align}
For a space $\mathcal{S}$ and matrix $X$, $\mathcal{P}_K$ holds the following property:
\begin{align}
\mathrm{span}\qty(\mathcal{P}_K(X)S)\in \mathrm{span}\qty(XS)\;.
\end{align}
Therefore, we prove that the space $\mathcal{S}^{(k\leq)}(\ket{\psi})$ is encompassed in $\mathbb{K}_k(V,\ket{\psi})$ as follows:
\begin{align}
\mathcal{S}^{(k)}(\ket{\psi})\in\mathrm{span}\qty(V^k \ket{\psi}) 
\Rightarrow \mathcal{S}^{(k\leq)}(\ket{\psi}) \in \mathbb{K}_k(V,\ket{\psi})\;.
\end{align}

Suppose that the perturbation $V$ can be expanded in the operation basis as follows:
\begin{align}
V=\sum_{i,j}v_{ij}\ket{\psi_i}\bra{\psi_j} \;.
\label{eq:perturbation}
\end{align}
From \cref{eq:perturbation}, we can define a graph $G_F$ consisting of nodes and edges corresponding to the indices of the diagonal and the non-zero off-diagonal elements of the perturbation $V$, respectively.
The $k$th power of $V$ is decomposed as follows:
\begin{align}
V^k=
&\sum_{\bm{l}_{i}^{(k)}}\qty{\qty(\prod_{(l,m) \in \bm{l}^{(k)}} v_{lm}) \ket{\psi_i}\bra{\psi_i}} \nonumber \\
&+\sum_{\bm{w}_{i\neq j}^{(k)}}\qty{\qty(\prod_{(l,m) \in \bm{w}^{(k)}} v_{lm}) \ket{\psi_i}\bra{\psi_j}}\;,
\label{eq:higher-order_interactions}
\end{align}
where $\bm{l}_{i}^{(k)}$ and $\bm{w}_{i\neq j}^{(k)}$ correspond to a length-$k$ loop around a node $i$ and a length-$k$ walk between nodes $i$ and $j$ on the graph $G_F$, respectively.
The graphical interpretation of the perturbation is sometimes called the path-sum approach~\cite{bravyi2011schrieffer, PhysRevResearch.2.023081}.
For verifying up to the $k$th-order Floquet collisions between a pair of Floquet states, we need to calculate up to the $k$th-order energy shifts and effective coupling strengths between them.
From \cref{eq:higher-order_interactions}, we can find that $k$th-order energy shifts and effective couplings correspond to the length-$k$ loop around the Floquet states and the length-$k$ walk between them, respectively.
As shown in \cref{fig:graphical_perturbation}~(a), such loops and walks are encompassed in the region within the distance $d=\lfloor{k/2\rfloor}$ from the both Floquet states.
Thus, to verify up to $k$th-order Floquet collisions caused on a Floquet state, we need to apply the same procedure for all Floquet states at distance $k$ from the Floquet state.
Such a process is encompassed in the region within distance $d=\lfloor{3k/2\rfloor}$ from the Floquet state on the Floquet subspace Hamiltonian.

Finally, we discuss the search and verification of all possible Floquet collisions for a given Floquet Hamiltonian.
The periodicity of the Floquet Hamiltonian indicates that the Floquet collisions that occur between $\ket{x,\va{n}_x}$-$\ket{y,\va{n}_y}$ also occur between $\ket{x,\va{n}_x+\va{r}}$-$\ket{y,\va{n}_y+\va{r}}$ for any translation operation $\va{r}$ to the $\mathrm{BZ}$.
Therefore, all Floquet collisions that occur in the Floquet Hamiltonian can be ascertained by verifying only the Floquet collisions that occur in the $\va{0}$th $\mathrm{BZ}$.

\subsection{Collision order and distance in real space}\label{sec:real_space}
In \cref{sec:floquet_space}, we laid out a correspondence between collision order and distance on the Floquet Hamiltonian.
For applications to quantum systems involving many qubits, it is also crucial to know the correspondence between collision order and distance in real space, in order to truncate our model down to the optimal complexity.

First, we note that Floquet states at distance $d$ do not always experience $d$th-order Floquet collisions.
As an example, suppose two isolated qubits individually irradiated by microwave drives as shown in \cref{fig:graphical_perturbation}~(b).
Despite no qubit-qubit interaction, there is a walk of distance $2$ between $\ket{0,0;0,0}$ and $\ket{1,1;-1,-1}$ mediated by $\ket{0,1;0,-1}$ and $\ket{1,0;-1,0}$ on the Floquet Hamiltonian.
However, they cannot collide with each other in any parameter regime, because the effective couplings between them caused by the two coupling walks always cancel each other out.

Such a relationship can be written more generally as follows. Floquet states are coupled to each other by off-diagonal terms in the Floquet Hamiltonian.
Each off-diagonal term corresponds to an interaction in real space according to \cref{eq:floquet_eigen_equation}.
Consider an $x$-body term in the interaction $H_I$ is acting on the qubits $\qty{Q_i}_{i=1}^{x}$.
The interaction then corresponds to node $\qty{Q_i}_{i=1}^{x}$ and the edge that joins them all.
Let $\qty{H_I}$ now be the pair of interactions that relay a particular walk between particular Floquet states on the Floquet Hamiltonian.
Consider a graph $G_R$ that consists of all the nodes and edges corresponding to the interactions $\qty{H_I}$.
If $G_R$ is not a connected graph, such a walk must be ignored because it will always cancel out with other similar walks.
In the following, such a walk is called an invalid walk and the converse is called a valid walk.
We will also refer to the qubits relevant to the nodes of the connected graph $G_R$ as the qubits involved in the walk.
	
Consider a qubit lattice where qubits couple via two-body interactions under strong-dispersive conditions.
A valid walk simultaneously involving qubits in a distance $d$ on the lattice must relay at least $d$ two-body interactions, and thus always has a length $d$ or more.
From the discussion in \cref{sec:floquet_space}, $k$th-order Floquet collisions caused on a Floquet state are complete within a subspace of distance $d=\lfloor{3k/2\rfloor}$ from the state.
Thus, we can say that the only qubits that can simultaneously be involved in a $k$th-order Floquet collisions involving a qubit are those within a distance $d=\lfloor{3k/2\rfloor}$ from that qubit.

\section{Collision analysis and computational complexity} \label{sec:complexity}
In \cref{sec:perturbation}, we have shown that for the analysis of Floquet collisions it is sufficient to work with a truncated finite dimensional Floquet subspace Hamiltonian in the $\va{0}$th $\mathrm{BZ}$.
However, the dimension of the $\va{0}$th $\mathrm{BZ}$ scales exponentially with the number of qubits, making its analysis computationally difficult.
In this section, we propose a collision analysis method that scales linearly with the number of qubits by considering the lattice structure of multi-qubit systems.
In the following, we consider the case where the operation basis can be expressed as a tensor product state for each qubit.
Note that we can also treat the cases with an entangling operation basis using \Cref{collision_finder_w_nlgates} in \cref{sec:truncated_sys}.

\subsection{Collision analysis for general qubit lattice}
\begin{figure*}
\begin{minipage}{\linewidth}
\begin{algorithm}[H]
\caption{Collision analysis for general qubit lattice} 
\label{collision_finder_wo_structure}
\begin{algorithmic}[1]
    \State Apply the discrete Fourier expansion to the target Hamiltonian as in \cref{eq:fourier_series_H_expansion}
    \State Construct a Floquet Hamiltonian from the expansion coefficients of the discrete Fourier series as \cref{eq:floquet_eigen_equation}
    \State Represent the Floquet Hamiltonian in the operation basis
    \State Convert the matrix representation of the Floquet Hamiltonian into a graph
    \State Around each node of interest in the graph, extract a subgraph within radius of $\lfloor{3k/2\rfloor}$.
    \State Perform $(k-1)$th-order diagonalization of the Floquet subspace Hamiltonian corresponding to the subgraph as \cref{eq:perturbative_diagonalization}
    \State Calculate the collision angle between all states of interest in the $\va{0}$th $\mathrm{BZ}$ relative to the other states in the $(k-1)$th-order diagonalized Hamiltonian as \cref{eq:floquet_quantization_angle}
    \State Identify the states with collision angles greater than a threshold as causing $k$th-order Floquet collisions
\end{algorithmic}
\end{algorithm}
\end{minipage}
\end{figure*}
Based on the discussions of the previous sections, we summarize our Floquet-based collision analysis for the general qubit lattice in \Cref{collision_finder_wo_structure}.
Consider the computational complexity $C_1(k,n,d,m)$ of $k$th-order collision analysis on a general $n$-qudit lattice, with all-to-all connectivity, where each qudit has $d$ levels and is irradiated with at most $m$ different microwave drive frequencies.
The corresponding $\va{0}$th $\mathrm{BZ}$ of the Floquet Hamiltonian has dimension $\order{d^{n}}$.
There are $\order{(m+n)^{\lfloor{3k/2\rfloor}}}$ nodes within $\lfloor{3k/2\rfloor}$ distance from a particular node in the Floquet Hamiltonian.
The computational complexity of the $k$th-order perturbative diagonalization of an $N\times N$-dimensional matrix is $\order{2^k N^3}$.
Therefore the total computational complexity $C_1(k,n,d,m)$ is given as follows:
\begin{align}
C_1(k,n,d,m)=\order{2^k\qty{d^n \qty(m+n)^{\lfloor{\frac{3k}{2}\rfloor}}}^3}\;.
\label{eq:computational_complexity_wo}
\end{align}
From \cref{eq:propagator_simple}, we can find that the gate errors caused by Floquet collisions depend mainly on the subspace spanned by the colliding Floquet states.
Therefore, the Floquet collisions of the computational Floquet states deteriorate the gate fidelity, and we can apply $d=2$ to \cref{eq:computational_complexity_wo}.
Because the dimension of the $\va{0}$th $\mathrm{BZ}$ increases exponentially with the number of qubits, \Cref{collision_finder_wo_structure} is feasible only for a limited number of qubits.

\subsection{Collision analysis for sparse qubit lattice} \label{sec:w_structure}
\begin{figure*}
\begin{minipage}{\linewidth}
\begin{algorithm}[H]
\caption{Collision analysis for sparse qubit lattice} 
\label{collision_finder_w_structure}
\begin{algorithmic}[1]
    \For {center qubit from the qubit lattice}
        \State Extract a sublattice including only the qubits within distance $\lfloor{3k/2\rfloor}$ from the center qubit
        \State Construct a subspace Hamiltonian $H^{\mathrm{sub}}$ corresponding to the sublattice
        \State Verify $k$th-order Floquet collisions involving the center qubit with \Cref{collision_finder_wo_structure}
        \State Remove the center qubit from the qubit lattice
    \EndFor
\end{algorithmic}
\end{algorithm}
\end{minipage}
\end{figure*}
Promising error correction codes, such as the surface code~\cite{gottesman1997stabilizer, bravyi1998quantum, fowler2012surface} and the color code~\cite{PhysRevLett.97.180501}, require qubits to be arranged in a periodic lattice structure.
The distance between qubits can be defined by taking the qubits as nodes and the nonzero exchange interaction between them as edges.
As discussed in \cref{sec:real_space}, only qubits within a distance $d=\lfloor{3k/2\rfloor}$ from a particular qubit can be involved in up to $k$th-order Floquet collisions affecting the qubit.
Therefore, we can verify all possible $k$th-order Floquet collisions by analyzing the corresponding $d=\lfloor{3k/2\rfloor}$ sublattice around each qubit sequentially.
Thus, the collision analysis for the sparse qubit lattice is summarized in \Cref{collision_finder_w_structure}.
Consider the computational complexity $C_2(k,n,d,m,r)$ of $k$th-order collision analysis on a sparse $n$-qudit lattice of maximum degree $r$, where each qudit has $d$ levels and is irradiated with at most $m$ different microwave drive frequencies.
We first select the center qudit and extract a sublattice containing $\order{r^{\lfloor{3k/2\rfloor}}}$ qudits each time.
The $\va{0}$th $\mathrm{BZ}$ of the Floquet Hamiltonian, corresponding to each sublattice, has dimension $\order{d^{r^{\lfloor{3k/2\rfloor}}}}$.
There are $\order{(m+r)^{\lfloor{3k/2\rfloor}}}$ nodes within $\lfloor{3k/2\rfloor}$ distance from a particular node on the Floquet Hamiltonian.
Therefore, the total computational complexity $C_2(k,n,d,m,r)$ is given as follows:
\begin{align}
C_2(k,n,d,m,r)=\order{n2^k\qty{d^{r^{\lfloor{\frac{3k}{2}\rfloor}}} \qty(m+r)^{\lfloor{\frac{3k}{2}\rfloor}}}^3}\;,
\label{eq:computational_complexity}
\end{align}
which is linear with respect to the number of qudits $n$.
Similar to \cref{eq:computational_complexity_wo}, we can apply $d\rightarrow 2$ when dealing only with fidelity deterioration caused by the Floquet collisions.

\section{Demonstration} \label{sec:demonstration}
In this section, we apply our Floquet-based collision analysis to relevant experimental systems and protocols.
First, we apply \Cref{collision_finder_wo_structure} to CR gates~\cite{rigetti2010fully, Sheldon_2016, Magesan_2020} on isolated two- or three-transmon systems, then calculate analytical solutions of the collision bounds which have been studied numerically so far~\cite{hertzberg2020laserannealing}.
Next, we apply \Cref{collision_finder_w_structure} to a heavy-hexagon code~\cite{PhysRevX.10.011022, neeraja2022matching} for quantitative estimation of the difficulty of system frequency allocation.

\subsection{Model}
We first describe our model of fixed-frequency transmon qubits~\cite{PhysRevA.76.042319, PhysRevB.77.180502} coupled via exchange interactions as follows:
\begin{align}
H=
& \sum_i 
\qty{
    \omega_i a_i^{\dag}a_i
    + \frac{\alpha_i}{2} a_i^{\dag}a_i^{\dag}a_i a_i
    } \nonumber \\
+
& \sum_{i,j}
\qty{
    J_{ij}
    \qty(a_i^\dagger + a_i)
    \qty(a_j^\dagger + a_j)
    } \nonumber \\
+
& \sum_k
\qty{
    \Omega_{k}\cos\qty(\omega_{dk}t+\phi_k) \qty(a_{t_k}^\dagger + a_{t_k})
    }\;,
\label{eq:system_hamiltonian}
\end{align}
where $\omega_i$, $\alpha_i$ and $J_{ij}$ are qubit frequencies, anharmonicities, and pairwise exchange interactions.
Moreover, $\Omega_k$, $\omega_{dk}$ and $\phi_k$ are the amplitude, frequency and phase of the $k$th-microwave drive on the target qubit $t_k$, respectively.
Annihilation operator for $i$th qubit is shown as $a_i$.

\subsection{Cross-resonance gate on an isolated-transmon system} \label{sec:demo_cr}
\begin{figure}[h!]
    \centering
	\includegraphics[width=0.45\textwidth]{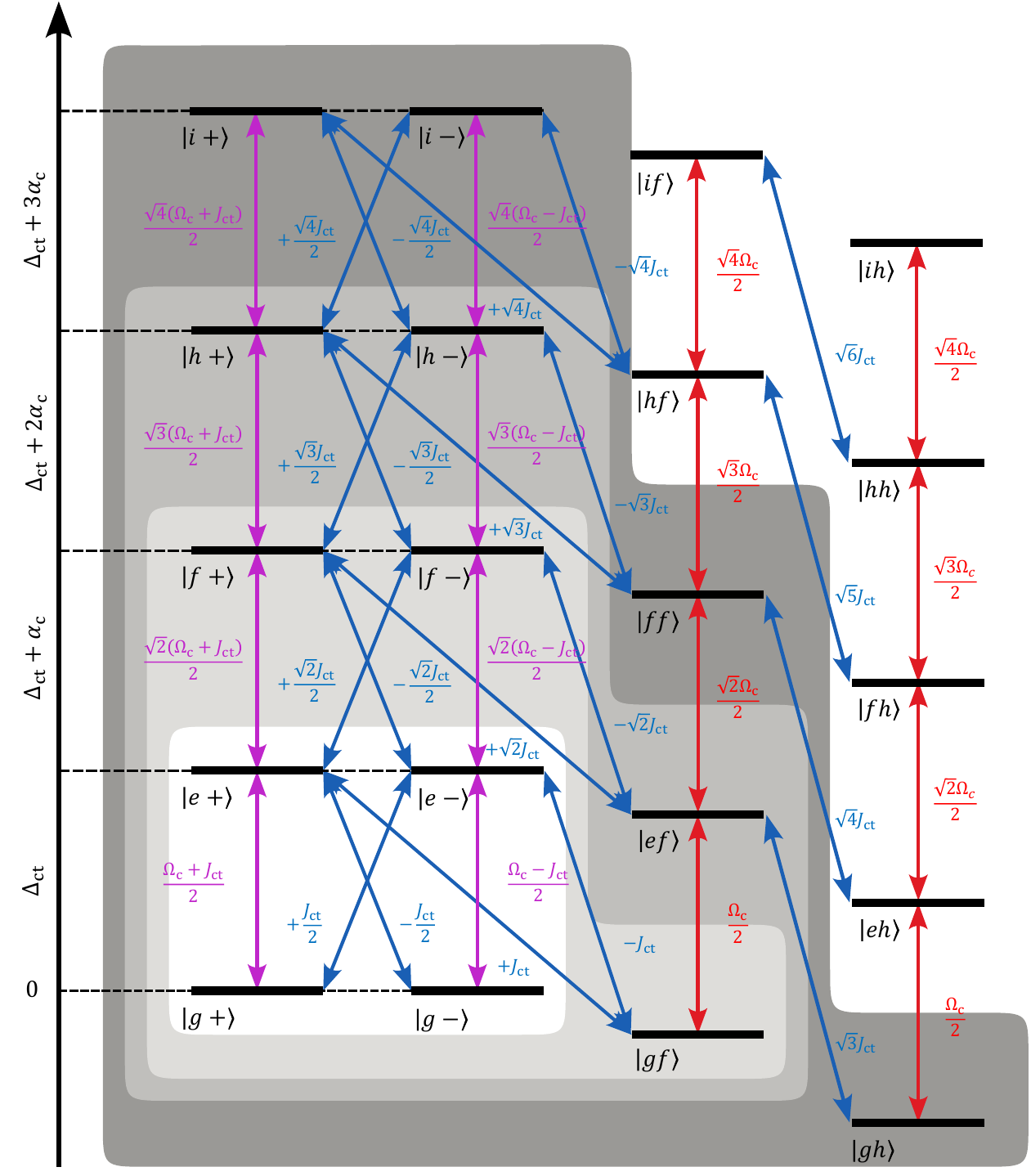}
    \caption{
    Floquet energy level diagram of the CR Hamiltonian~(\cref{eq:cr_hamiltonian}).
    For simplicity, the CR drive frequency is set to the bare target qubit frequency $\omega_t$ .
    Also, the $\mathrm{BZ}$ index of the Floquet states is omitted.
    The gray regions represent the distance from the computational states $\qty{\ket{g\pm},\ket{e\pm}}$, respectively, with darker shades further away.
    Although the $\ket{x\pm}$ states appear to be degenerate for $x\in[g,e,f,h,i\cdots]$, they are in fact detuned by a second-order energy shift originating from the CR interaction, preserving the strong-dispersive condition.
    }
    \label{fig:demo_cr}
\end{figure}
In this subsection, we apply \Cref{collision_finder_wo_structure} to Floquet collisions of CR gates on isolated few-transmon systems, then present analytical solutions up to second-order in perturbation.
We first consider an isolated CR gate consisting of control and target qubits.
Next, we present a generalization with a third spectator qubit~\cite{Malekakhlagh_2020}.

To implement the CR gate, the control qubit is irradiated with a microwave drive resonant with the dressed frequency of the target qubit.
Following \cref{eq:system_hamiltonian}, the system Hamiltonian is
\begin{align}
H=
& \sum_{i\in[c,t]}
\qty{
    \omega_ia_i^{\dag}a_i
    + \frac{\alpha_i}{2} a_i^{\dag}a_i^{\dag}a_i a_i
    } \nonumber \\
& +
J_{ct}
\qty(a_c^\dagger + a_c)
\qty(a_t^\dagger + a_t)  \nonumber \\
& +
\Omega_{c}\cos\qty(\tilde{\omega}_{t}t) \qty(a_{c}^\dagger + a_{c})\;,
\label{eq:cr_hamiltonian}
\end{align}
with indices ``c'' and ``t'' referring to the control and the target qubits, respecitively.
The drive frequency should be set to the \textit{dressed} target frequency $\tilde{\omega}_t$ found perturbatively up to $O(J_{ct}^4)$ as
\begin{align}
\tilde{\omega}_t \approx \omega_t - \frac{J_{ct}^2}{\Delta_{ct}} + 
\frac{(\alpha_c + \alpha_t)J_{ct}^2}{(\Delta_{ct}+\alpha_c)(\Delta_{ct} - \alpha_t)}\;,
\end{align}
where $\Delta_{ct}\equiv\omega_{c}-\omega_{t}$.
Moreover, for the spectator problem, we consider a third transmon qubit, coupled to the control, denoted by index $s$.

Under such a CR drive scheme, a controlled-$X$ rotation occurs in the computational subspace of the target qubit depending on the state of the control qubit.
We therefore choose the Floquet operation basis as:
\begin{align}
\ket{c,t;n}&=\ket{c}\otimes\ket{t;n} \;,\\
\ket{\pm;n}&=\frac{1}{\sqrt{2}}\qty(\ket{g;n}\pm\ket{e;n-1})\;,
\label{eq:cr_operation}
\end{align}
with $c \in [g,e,f,\cdots]$, $t \in [+,-,f,\cdots]$, and the $\mathrm{BZ}$ index $n \in \mathbb{Z}$.
In the following, we omit $\mathrm{BZ}$ index $n$ for simplicity.

\cref{fig:demo_cr} shows the Floquet energy level diagram in the operation basis.
Using \Cref{collision_finder_wo_structure}, we verify up to second-order Floquet collisions involving the computational basis $\qty{\ket{g,\pm}, \ket{e,\pm}}$.
First, we extract the subgraph consisting of nodes within a distance of $3$ from the computational subspace.
Next, we apply perturbative diagonalization on a Floquet subspace Hamiltonian reconstructed from the subgraph.
Finally, we calculate collision factors: fractions of the off-diagonal terms to the detuning between the diagonal terms as linearized collision angles between the corresponding Floquet states.
For simplicity, we omitted terms $\order{J_{ct}^x \Omega ^y}$ for $x+y \geq 3$ in the following.

\renewcommand{\arraystretch}{2.5}
\begin{table*}
	\caption{
	Analytical solution of Floquet collisions for the two-transmon CR system up to second-order in perturbation.
    The first column represents the degenerate Floquet state pairs in the operation basis, where we omitted the $\mathrm{BZ}$ index for simplicity.
    The second column shows the labels for the frequency collisions under which the collision factors of the corresponding Floquet collisions diverge as in \Cref{tab:collision_freq}.
    From the second column, we find that multiple Floquet state pairs can become degenerate for a given frequency condition.
    The third column shows the order of the Floquet collisions defined in \cref{sec:floquet_space}.
    The fourth column shows the collision factors, i.e. the linearized collision angles.
    We can verify Floquet collisions by checking whether the collision factors are sufficiently smaller than $1$.
	}
	\label{tab:collision_2t}
	\begin{tabular}{p{2cm}p{2cm}p{1cm}p{12cm}} \hline \hline
	    $\ket{c,t}$                 & Type  & Order & Collision Factor \\
	    \hline
		$\ket{g\pm}\leftrightarrow\ket{e\pm}$     & 1     & 1     & $\displaystyle \frac{J_{ct}\pm\Omega_{c}}{\Delta_{ct}}$   \\ 
		$\ket{g\pm}\leftrightarrow\ket{e\mp}$     & 1     & 1     & $\displaystyle \frac{J_{ct}}{\Delta_{ct}}$                 \\
		$\ket{gf}\leftrightarrow\ket{e\pm}$       & 3     & 1     & $\displaystyle \frac{J_{ct}}{\Delta_{ct}-\alpha_{t}}$       \\  
		$\ket{e\pm}\leftrightarrow\ket{f\pm}$     & 3     & 1     & $\displaystyle \frac{J_{ct}\pm\Omega_{c}}{\sqrt{2}\qty(\Delta_{ct}+\alpha_c)}$      \\  
		$\ket{e\pm}\leftrightarrow\ket{f\mp}$     & 3     & 1     & $\displaystyle \frac{J_{ct}}{\sqrt{2}\qty(\Delta_{ct}+\alpha_c)}$                    \\ 
		$\ket{g+}\leftrightarrow\ket{g-}$         & 3     & 2     & $\displaystyle \frac{\Delta_{ct}J_{ct}(\alpha_{c}+\alpha_{t})}{2\Omega_{c}(\Delta_{ct}+\alpha_{c})(\Delta_{ct}-\alpha_{t})}$     \\
		$\ket{g\pm}\leftrightarrow\ket{gf}$         & 1,3     & 2     & $\displaystyle \pm\frac{\Omega_{c}J_{ct}}{4\alpha_{t}} \qty(\frac{1}{\Delta_{ct}-\alpha_{t}} + \frac{1}{\Delta_{ct}})$                 \\
		$\ket{g\pm}\leftrightarrow\ket{f\pm}$         & 1,2,3   & 2     & $\displaystyle \frac{\sqrt{2}\Omega_{c}\alpha_{c}(\Omega_{c}\pm2J_{ct})}{8\Delta_{ct}(\Delta_{ct}+\alpha_{c})(2\Delta_{ct}+\alpha_{c})}$ \\
		$\ket{g\pm}\leftrightarrow\ket{f\mp}$         & 1,2,3   & 2     & $\displaystyle \pm\frac{\sqrt{2}\Omega_{c}\alpha_{c}J_{ct}}{4\Delta_{ct}(\Delta_{ct}+\alpha_{c})(2\Delta_{ct}+\alpha_{c})}$    \\
		$\ket{e+}\leftrightarrow\ket{e-}$         & 3,8     & 2     & $\displaystyle \frac{\Delta_{ct}J_{ct}(\alpha_{c}+\alpha_{t})}{2\Omega_{c}(\Delta_{ct}-\alpha_{c})(\Delta_{ct}-\alpha_{t})}$   \\
		$\ket{e\pm}\leftrightarrow\ket{ef}$         & 1,3,9     & 2     & $\displaystyle
		\pm\frac{\Omega_{c}J_{ct}}{2\alpha_{t}}\qty(\frac{1}{\Delta_{ct}+\alpha_{c}}+\frac{1}{\Delta_{ct}+\alpha_{c}-\alpha_{t}}-\frac{2\Delta_{ct}-\alpha_{t}}{2\Delta_{ct}(\Delta_{ct}-\alpha_{t})})$\\
		$\ket{e\pm}\leftrightarrow\ket{h\pm}$         & 2,3,10,11     & 2     & $\displaystyle \frac{\sqrt{6} \Omega_{c} \alpha_{c} (\Omega_{c}\pm2J_{ct})}{8(\Delta_{ct}+\alpha_{c})(\Delta_{ct}+2\alpha_{c})(2\Delta_{ct}+3\alpha_{c})} $\\
		$\ket{e\pm}\leftrightarrow\ket{h\mp}$         & 2,3,10,11     & 2     & $\displaystyle \pm\frac{\sqrt{6}\Omega_{c}\alpha_{c}J_{ct}}{4 (\Delta_{ct}+\alpha_{c})(\Delta_{ct}+2\alpha_{c})(2\Delta_{ct}+3\alpha_{c})} $\\
		\hline \hline
	\end{tabular}
\end{table*}
\Cref{tab:collision_2t} shows the pairs of Floquet states involved in up to second-order Floquet collisions and their corresponding collision factors.
Here, from the discussion in \cref{sec:complexity}, we chose only the pairs with at least one computational state.
\renewcommand{\arraystretch}{2.5}
\begin{table*}
	\caption{
	Analytical solution of Floquet collisions for  the three-transmon CR system with a control spectator qubit up to second order.
    We show the Floquet state pairs, corresponding frequency collisions, collision orders, and collision factors for various Floquet collisions similar to \Cref{tab:collision_2t}.
	}
	\label{tab:collision_3t}
	\begin{tabular}{p{3cm}p{2cm}p{1cm}p{12cm}} \hline \hline
	    $\ket{c,t,s}$                  & Type  & Order & Collision Factor \\
	    \hline
		$\ket{g\pm e}\leftrightarrow\ket{e\pm g}$     & 5     & 1     & $\displaystyle -\frac{J_{cs}}{\Delta_{cs}}$   \\ 
        $\ket{g\pm f}\leftrightarrow\ket{e\pm e}$     & 6     & 1     & $\displaystyle \frac{\sqrt{2}J_{cs}}{\alpha_{s}-\Delta_{cs}}$   \\ 
        $\ket{e\pm e}\leftrightarrow\ket{f\pm g}$     & 6     & 1     & $\displaystyle -\frac{\sqrt{2}J_{cs}}{\alpha_{c}+\Delta_{cs}}$   \\ 
        $\ket{g\pm g}\leftrightarrow\ket{g\pm e}$     & 1,5,12     & 2     & $\displaystyle \frac{\Omega_{c}J_{cs}}{4\Delta_{st}}\qty(\frac{1}{\Delta_{ct}}+\frac{1}{\Delta_{cs}})$   \\ 
        $\ket{g\pm e}\leftrightarrow\ket{g\pm f}$     & 1,6,13     & 2     & $\displaystyle \frac{\sqrt{2}\Omega_{c}J_{cs}}{\Delta_{st}+\alpha_{s}} \qty(\frac{1}{\Delta_{ct}}-\frac{1}{(\Delta_{sc}+\alpha_{s})})$   \\
        $\ket{g\pm e}\leftrightarrow\ket{f\pm g}$     & 1,3,5,6,7     & 2     & $\displaystyle \frac{\sqrt{2}\Omega_{c}\alpha_{c}J_{cs}}{4(\Delta_{ct}+\Delta_{cs}+\alpha_{c})}\qty(\frac{1}{\Delta_{ct}(\Delta_{ct}+\alpha_c)}+\frac{1}{\Delta_{cs}(\Delta_{cs}+\alpha_{c})})$   \\
        $\ket{e\pm g}\leftrightarrow\ket{e\pm e}$     & 1,5,3,6,12    & 2     & $\displaystyle \frac{\Omega_{c} J_{cs}}{4\Delta_{st}} \qty(\frac{2}{\Delta_{cs} + \alpha_{c}} + \frac{2}{\Delta_{ct} + \alpha_{c}} - \frac{1}{\Delta_{cs}} - \frac{1}{\Delta_{ct}})$   \\
        $\ket{e\pm e}\leftrightarrow\ket{e\pm f}$     & 1,3,6,13,14     & 2     & $\displaystyle \sqrt{2}\Omega_{c}J_{cs} \left(
        \frac{1}{(\Delta_{st}+\alpha_{s})(\Delta_{cs}+\alpha_{c}-\alpha_{s})}
        - \frac{1}{4(\Delta_{cs}-\alpha_{s})(\Delta_{st}+\alpha_{s})}\right.$ \\
        &   &   & $\displaystyle \left. - \frac{1}{2(\Delta_{ct}+\alpha_{c})(\Delta_{cs}+\alpha_{c}-\alpha_{s})}
        - \frac{1}{4\Delta_{ct}(\Delta_{st}+\alpha_{s})}\right)$ \\
        $\ket{e\pm e}\leftrightarrow\ket{h\pm g}$     & 3,4,6,14,15     & 2     & $\displaystyle \frac{\sqrt{6}\Omega_{c}\alpha_{c}J_{cs}}{4(\Delta_{ct}+\Delta_{cs}+3\alpha_{c})}\qty(\frac{1}{(\Delta_{ct}+\alpha_{c})(\Delta_{ct}+2\alpha_{c})}+\frac{1}{(\Delta_{cs}+\alpha_{c})(\Delta_{cs}+2\alpha_{c})})$   \\
		\hline \hline
	\end{tabular}
\end{table*}
We also calculate the case with a spectator qubit coupled to the control qubit and show the results in \Cref{tab:collision_3t}, where $\omega_{s}$ and $J_{cs}$ are the spectator qubit frequency and the control-spectator coupling strength, respectively.
We only show Floquet collisions with the factor of $\order{J_{ct}^x J_{cs}^y \Omega ^z}$ with $x+y < 2$ and $z < 3$ for brevity, but in principle our analysis can provide further higher-order Floquet collisions.
\renewcommand{\arraystretch}{1}
\begin{table}
	\caption{
	Analytical solutions of the frequency collisions where the corresponding collision factors cited in \Cref{tab:collision_2t} and \Cref{tab:collision_3t} diverges.
    The labels used here follow the previous empirical classification of the frequency collisions~\cite{hertzberg2020laserannealing, doi:10.1126/sciadv.abi6690}.
    We note, however, that the previously labeled type 4 collision does not correspond to an actual frequency collision under our Floquet-based collision analysis.
    It is an empirical frequency allocation requirement to have control-target detuning in the straddling regime to have stronger CR interaction.
    On the other hand, type 8 is a novel frequency collision which has not been reported in the previous studies.
    Type 8 is caused by the mutual cancellation of entanglement interaction and quantum crosstalk in the CR drive.
    Detailed discussion is found in \cref{sec:demo_cr}.
	}
	\label{tab:collision_freq}
	\begin{tabular}{p{1cm}p{5cm}p{1cm}} \hline \hline
	    Type & Frequency condition & Order \\
	    \hline
        1   & $\displaystyle \Delta_{ct}=0$ & 1 \\
        2   & $\displaystyle 2\Delta_{ct}+\alpha_{c}=0$ & 2 \\
        3   & $\displaystyle \Delta_{ct}+\alpha_{c}=0~\mathrm{or}~\Delta_{tc}+\alpha_{t}=0$ & 1 \\
        5   & $\displaystyle \Delta_{cs}=0$ & 1 \\
        6   & $\displaystyle \Delta_{cs}+\alpha_{c}=0~\mathrm{or}~\Delta_{sc}+\alpha_{s}=0$ & 1 \\
        7   & $\displaystyle \Delta_{ct}+\Delta_{cs}+\alpha_{c}=0$ & 2 \\
        8   & $\displaystyle \Delta_{ct}-\alpha_{c}=0$ & 2 \\
        9   & $\displaystyle \Delta_{ct}+\alpha_{c}-\alpha_{t}=0$ & 2 \\
        10  & $\displaystyle \Delta_{ct}+2\alpha_{c}=0$ & 2 \\
        11  & $\displaystyle 2\Delta_{ct}+3\alpha_{c}=0$ & 2 \\
        12  & $\displaystyle \Delta_{st}=0$ & 2 \\
        13  & $\displaystyle \Delta_{st}+\alpha_{s}=0~\mathrm{or}~\Delta_{st}-\alpha_{t}=0$ & 2 \\
        14  & $\displaystyle \Delta_{ct}+\Delta_{cs}+3\alpha_{c}=0$ & 2 \\
        15  & $\displaystyle \Delta_{cs}+2\alpha_{c}=0$ & 2 \\
		\hline \hline
	\end{tabular}
\end{table}
From \Cref{tab:collision_2t} and \Cref{tab:collision_3t}, we find several frequency conditions under which some of the collision factors diverge.
They correspond to the conventional frequency collisions~\cite{Malekakhlagh_2020,hertzberg2020laserannealing, doi:10.1126/sciadv.abi6690}.
We summarized the frequency collisions in \Cref{tab:collision_freq}.

Tables~\ref{tab:collision_2t} and \ref{tab:collision_3t} also show for which type of frequency collisions the collision factors diverge and the corresponding perturbation order in the effective coupling strength between Floquet states.
The collision factor diverges when the strong-dispersive condition is completely broken down, i.e. two of the Floquet states are precisely degenerate, in the valid walk between the target Floquet states on the Floquet Hamiltonian.
Thus, as shown in Tables~\ref{tab:collision_2t} and \ref{tab:collision_3t}, the first-order Floquet collisions with no mediating Floquet states have only one frequency collision, whereas the second-order Floquet collisions with multiple mediating Floquet states have multiple frequency collisions.

The frequency collisions in \Cref{tab:collision_freq} agree mostly with the results in the previous studies~\cite{Malekakhlagh_2020, hertzberg2020laserannealing}.
The differences from the previous studies are type 4 and type 8.
Type 4 was proposed in Refs.~\cite{hertzberg2020laserannealing, doi:10.1126/sciadv.abi6690} as an empirical straddling regime frequency allocation requirement for the coupled-transmon systems to have stronger CR interaction.
We, however, note that the type 4 collision does not correspond to an actual frequency collision under our Floquet-based collision analysis.
On the other hand, type 8 is a new frequency collision not mentioned previously~\cite{Malekakhlagh_2020, hertzberg2020laserannealing, doi:10.1126/sciadv.abi6690}.
As shown in \Cref{tab:collision_2t}, type 8 appears only in the Floquet collision between $\ket{e+}$ and $\ket{e-}$ states, which can be understood in terms of
the effective CR Hamiltonian ~\cite{Magesan_2020}:
\begin{align}
H_{\mathrm{CR}}&=\Omega_{ZX}\frac{ZX}{2}+\Omega_{IX}\frac{IX}{2}\;.
\label{eq:simple_cr_eff_h}
\end{align}
The energy detuning between the $\ket{e+}$ and $\ket{e-}$ states is expressed as $\Omega_{IX}-\Omega_{ZX}\propto \Delta_{ct}-\alpha_{c}$, and thus the collision factor is inversely proportional to $\Delta_{ct} - \alpha_c$.

From our perturbative estimates for the collision factors, we can also provide analytical bounds on fidelity degradation for a given Floquet collision as \cref{sec:floquet_collision}.
Our analytical collision factors would provide more flexible and precise system frequency allocation.
From Tables~\ref{tab:collision_2t} and \ref{tab:collision_3t}, we find that the collision factors depend not only on the qubit frequency, anharmonicity, and the coupling strength, but also on the microwave drive amplitude.
It suggests drive-induced Floquet collisions and can explain why the CR gate fidelity and execution time is limited in the previous experiments~\cite{jurcevic2021demonstration, PRXQuantum.2.040336}.
More rigorous analysis on adiabatic conditions for the Floquet collisions would enable faster CR gates.

The discussion so far provides analytical formulation of (i)~mechanism, (ii)~type, and (iii)~bounds of frequency collisions in two/three-qubit CR systems.
We can also perform numerical Floquet simulations to visualize the Floquet collisions.
\begin{figure}
    \centering
	\includegraphics[width=0.4\textwidth]{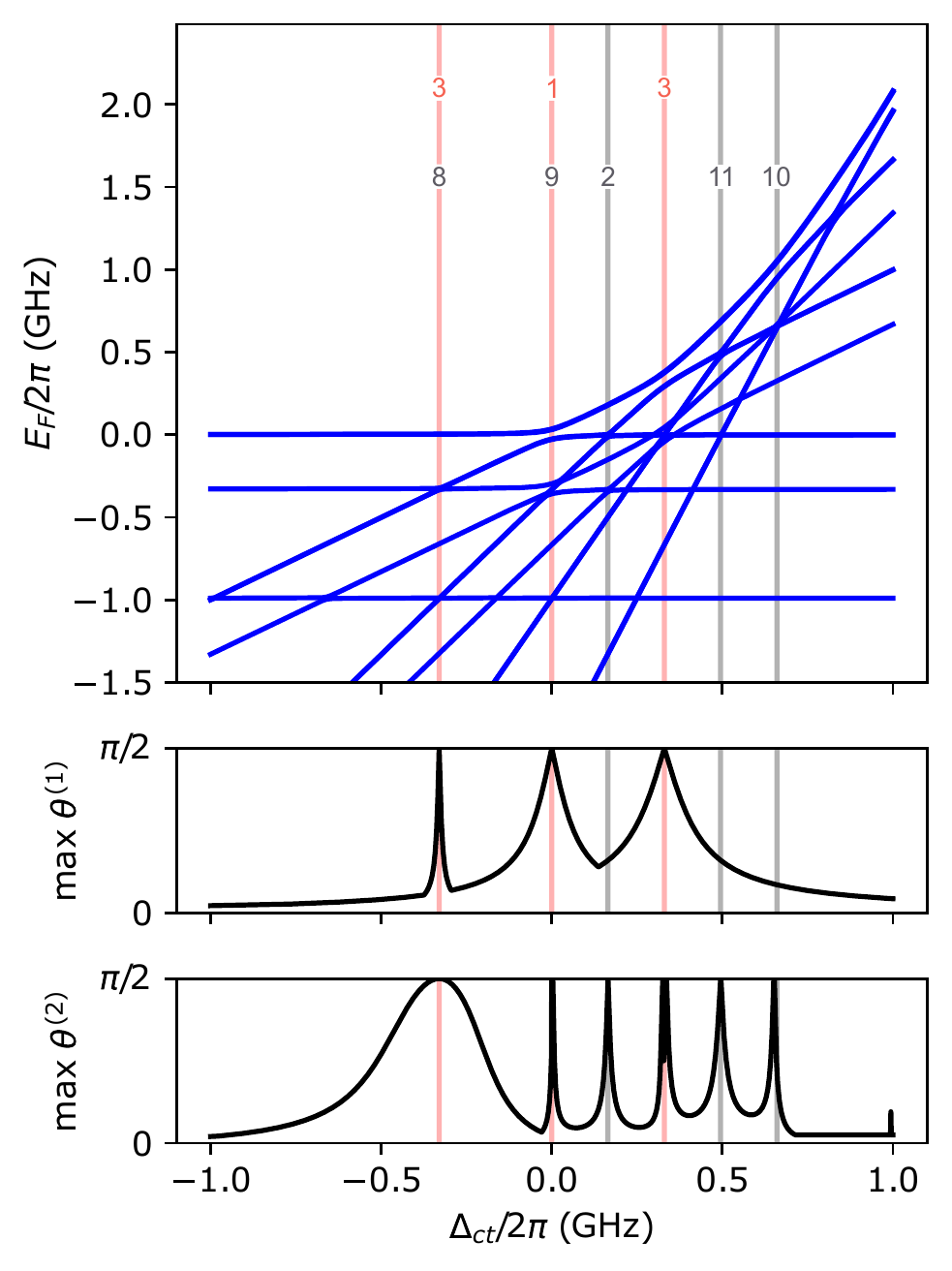}
    \caption{
    Numerical simulations of \Cref{collision_finder_wo_structure} for a two-qubit CR system while sweeping the control-target detuning $\Delta_{ct}/2\pi$ from $-1$ to $1$~GHz.
    Qubits anharmonicity and qubit-qubit coupling are set to $-330$ MHz and $3.8$ MHz, respectively.
    The top panel shows the Floquet quasi-excitation energies in the Floquet subspace generated for searching up to the second-order Floquet collisions.
    As a function of detuning, certain Floquet quasi-excitation energies shift and anti-cross with each other.
    The middle and lower panels represent the maximum collision angles between the Floquet state pairs corresponding to first- and second-order Floquet collisions, respectively.
    The vertical lines and their labels represent the frequency collisions shown in \Cref{tab:collision_freq}.
    }
    \label{fig:fqe_cr1}
\end{figure}
\Cref{fig:fqe_cr1} shows the simulation results of \Cref{collision_finder_wo_structure} for a two-qubit CR gate with fixed CR drive amplitude of $\Omega_c/2\pi = 30$~MHz while sweeping the control-target detuning $\Delta_{ct}/2\pi$ between $-1$ to $1$~GHz.
The top panel shows the Floquet quasi-excitation energies of the system.
As shown in the figure, some of the Floquet quasi-excitation energies change linearly with the detuning sweep and exhibit avoided crossings that we refer to as Floquet collisions.
The second and third rows of the figure show the maximum values of the collision angles corresponding to first- and second-order Floquet collisions, respectively.
Note that we plotted collision angles only for Floquet collisions involving the computational subspace.
The red and black vertical lines label the underlying first- and second-order frequency collisions, respectively.
Under the first-order frequency collisions, we find type 1 and type 3.
Among the type 3 cases, the case where the $ef$-transition frequency of the control qubit is degenerate to the $ge$-transition frequency of the target qubit has larger collision bound than the opposite case.
This is because the former corresponds to $\ket{e\pm}$-$\ket{f\pm}$ Floquet collision whose collision factor is proportional to $\Omega_c$, while the latter corresponds to a $\ket{gf}$-$\ket{e\pm}$ Floquet collision whose collision factor is proportional to $J_{ct}$.
Under the second-order frequency collisions, we find types 2,8,9,10,11 in addition.

\begin{figure*}
    \centering
	\includegraphics[width=\textwidth]{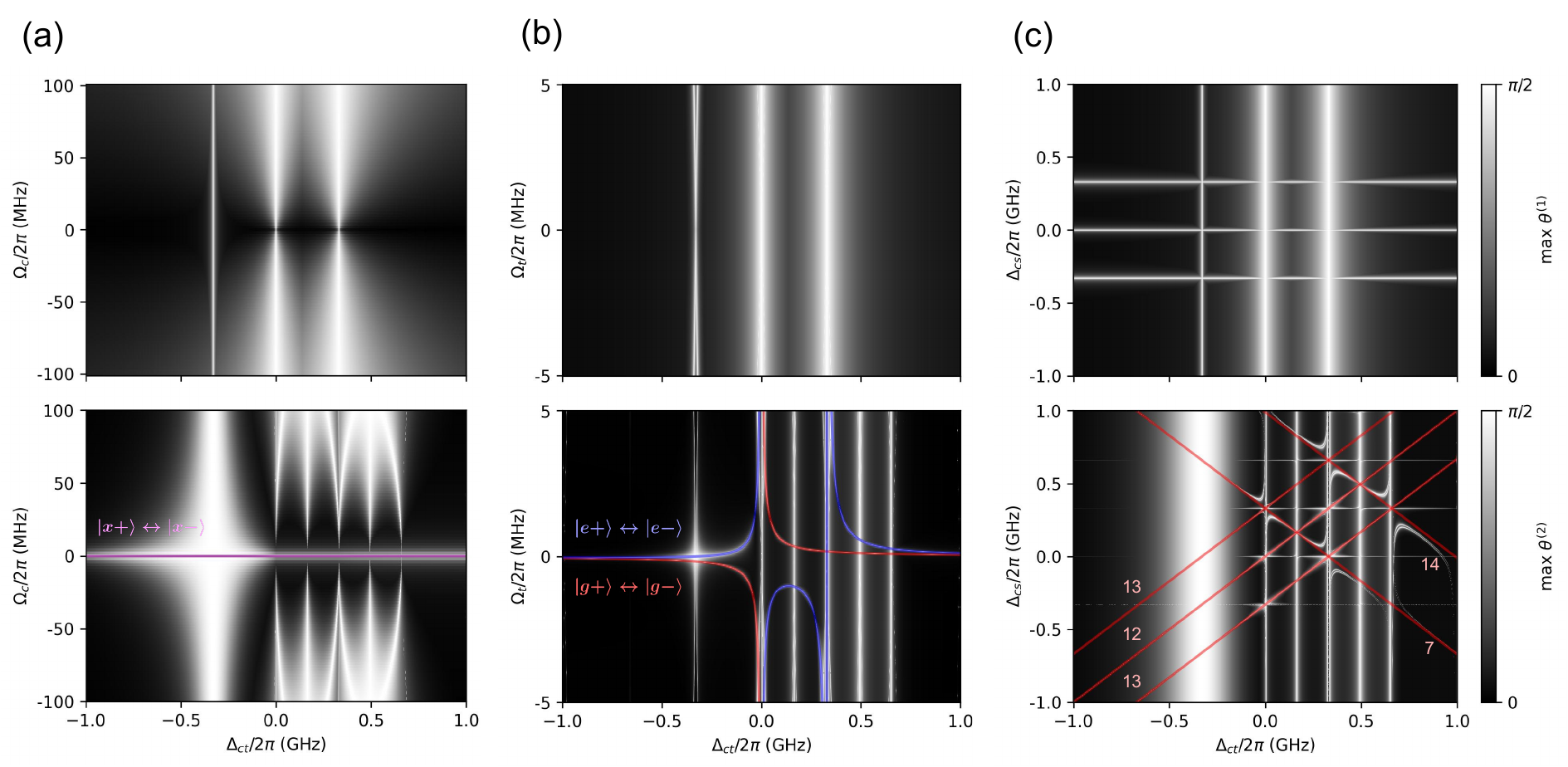}
    \caption{
    Numerical simulations of \Cref{collision_finder_wo_structure} for a CR drive on two~(a,b) or three~(c) transmon system with 2D sweep of system parameters.
    One of the sweeping system parameters is always control-target detuning $\Delta_{ct}/2\pi$ from $-1$ to $1$~GHz.
    Qubit anharmonicities qubit-qubit coupling strength is set to $-330$ MHz and $3.8$ MHz, respectively.
    The top~(bottom) panels show the gray-scale plot of first-order~(second-order) Floquet collision angles, respectively.
    The coloured lines and their labels represent the corresponding frequency or Floquet collisions.
    (a) sweeping the CR drive amplitude $\Omega_c/2\pi$ from $-100$ to $100$~MHz.
    (b) sweeping the rotary tone drive amplitude $\Omega_t/2\pi$ from $-5$ to $5$~MHz.
    (c) sweeping the frequency detuning between control and spectator qubits $\Delta_{cs}/2\pi$ from $-1$ to $1$~GHz.
    }
    \label{fig:fqe_cr2}
\end{figure*}
\Cref{fig:fqe_cr2} shows the simulation results of \Cref{collision_finder_wo_structure} for a CR drive on two~(a,b) or three~(c) qubits with 2D sweep of system parameters.
One of the sweeping system parameters is always the control-target detuning $\Delta_{ct}/2\pi$ from $-1$ to $1$~GHz.
In each figure, the top and bottom figures show gray-scale plots of the maximum collision angles corresponding to first- and second-order Floquet collisions, respectively.
The coloured lines and their labels represent the corresponding frequency or Floquet collisions.

In \cref{fig:fqe_cr2}~(a), we sweep the CR drive amplitude $\Omega_c/2\pi$ from $-100$ to $100$~MHz in addition to the control-target detuning.
First-order collision angle~(top panel) reveals one type 1 and two type 3 frequency collisions the same as \cref{fig:fqe_cr1}.
The width of the type 1~($\Delta_{ct}=0$) and one of the type 3 frequency collisions~($\Delta_{ct}+\alpha_c=0$) increases with stronger drive amplitude in agreement with \Cref{tab:collision_2t}, where the collision factors are proportional to $\Omega_c$.
The other type 3 frequency collision~($\Delta_{ct}-\alpha_t=0$), corresponding to the Floquet collision $\ket{gf} \leftrightarrow\ket{e\pm}$, is independent of the CR drive amplitude.
Second-order collision angles~(bottom panel) exhibit both increase/decrease in the linewidth as well as a Stark shift in frequency collisions with increasing CR drive amplitude.
Decreasing linewidth is observed only for type 8, i.e. Floquet collisions between $\ket{g+}$~($\ket{e+}$) and $\ket{g-}$~($\ket{e-}$), where the collision factor in \Cref{tab:collision_2t} is inversely proportional to $\Omega_c$, and therefore occurs independently of $\Delta_{ct}$ at $\Omega_c=0$.
This is due to the static-$ZZ$ interaction, causing the frequency of the target qubit to depend on the control qubit.
When the CR drive amplitude is $0$, in the operation basis, $\ket{g+}$~($\ket{e+}$) and $\ket{g-}$~($\ket{e-}$) have no energy gap caused by the CR drive and are therefore degenerate.

In \cref{fig:fqe_cr2}~(b), we add a rotary tone drive~\cite{Sheldon_2016, PRXQuantum.1.020318} into the system and sweep its amplitude $\Omega_t/2\pi$ from $-5$ to $5$~MHz.
Rotary tone is a kind of spin locking technique, which is a resonant drive to the target qubit employed to eliminate unwanted error terms having an anti-commutative relation to $IX$ in the effective CR Hamiltonian.
The top panel shows that the frequency collision, corresponding to the Floquet collision between $\ket{gf}$ and $\ket{e\pm}$, splits with increasing $\qty|\Omega_t|$, as a result of drive-induced Stark shift of the Floquet quasi-excitation energies of states $\ket{e\pm}$.
The bottom panel shows that the two peaks in the collision angles, corresponding to Floquet collisions between $\ket{g+}$~($\ket{e+}$) and $\ket{g-}$~($\ket{e-}$), are strongly dependent on $\Omega_t$.
The rotary tone causes the Floquet quasi-excitation energies of $\ket{g\pm}$ and $\ket{e\pm}$ to shift, where the detuning obeys:
\begin{align}
\Delta_{g\pm}&=\Omega_{t} + \Omega_{IX} + \Omega_{ZX}\;, \\
\Delta_{e\pm}&=\Omega_{t} + \Omega_{IX} - \Omega_{ZX}\;,
\end{align}
with $\Omega_{ZX}$ and $\Omega_{IX}$ being the $ZX$ and $IX$ term in the CR effective Hamiltonian~(\cref{eq:simple_cr_eff_h}).
In the bottom panel, the red and blue curves correspond to $\Delta_{g\pm}=0$ and $\Delta_{e\pm}=0$, respectively, which are well aligned with the peaks.
Therefore, our Floquet analysis clarifies the role of the rotary tone as a technique to artificially widen the detuning between $\ket{g+}$~($\ket{e+}$) and $\ket{g-}$~($\ket{e-}$), which suppresses the type 8 collision due to static-$ZZ$ interaction.

In \cref{fig:fqe_cr2}~(c), we introduce a spectator qubit, coupled only to the control qubit, and sweep the control-spectator detuning $\Delta_{cs}/2\pi$ from $-1$ to $+1$~GHz.
In the top panel, the vertical and horizontal peaks correspond to independent first-order Floquet collisions between the control-spectator and control-target qubits, respectively.
This is the case as first-order Floquet collisions only involve nearest-neighboring qubits following \cref{sec:real_space}.
The bottom panel, however, shows diagonal peaks corresponding to the frequency collisions between the target and spectator qubits,
allowed in second-order Floquet collisions.
Moreover, the red diagonal lines corresponding to type 7, 12, 13 and 14, which are well aligned with the diagonal peaks.
An interesting observation is that the peaks anti-cross each other at the intersections, which we refer to as ``collision avoided crossing''.
As discussed in \cref{sec:floquet_collision}, when multiple Floquet collisions involve the same Floquet state simultaneously, the collision conditions are derived by diagonalizing the subspace spanned by the colliding Floquet states.
The frequency collisions shown in \cref{tab:collision_freq} do not assume the simultaneous Floquet collisions, so the red diagonal lines overlook the collision avoided crossings.
These observations suggest that frequency allocation for large-scale quantum processors requires more involved analysis compared to the previous plans derived based on small-scale system considerations ~\cite{hertzberg2020laserannealing, doi:10.1126/sciadv.abi6690}.

\subsection{Syndrome extraction on a heavy-hexagon lattice}\label{sec:demo_qec}
\begin{figure*}
    \centering
	\includegraphics[width=\textwidth]{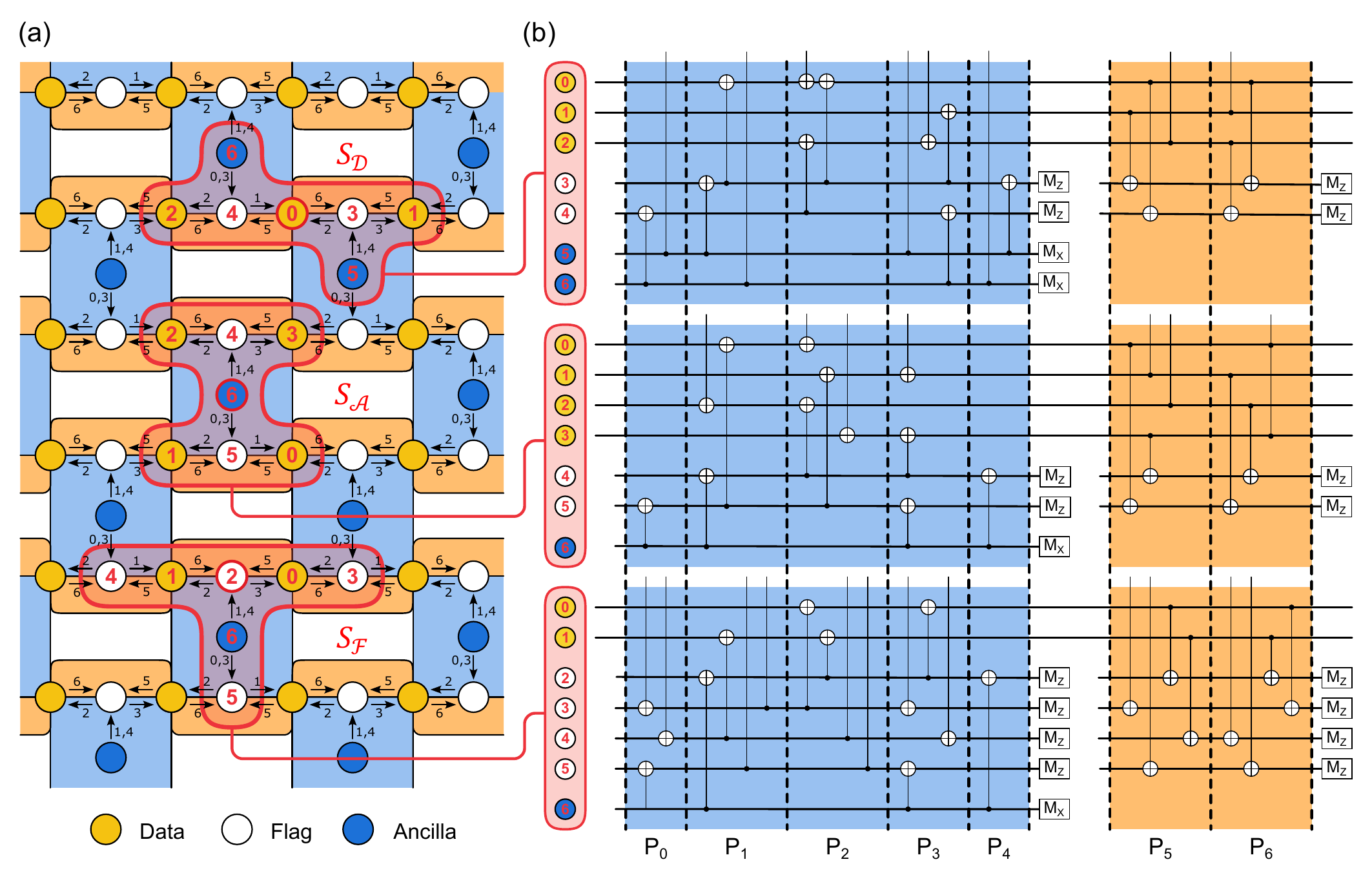}
    \caption{
    Procedures of the error syndrome extraction for the heavy-hexagon code~\cite{PhysRevX.10.011022}, consisting of seven types of simultaneous CR drives labeled $\qty{P_i}_{i=0}^{6}$.
    Yellow, blue, or white circles correspond to the data, ancilla, and flag qubits, respectively.
    Blue and Orange tiles represent the $Z$- and $X$-type parity measurements.
    Red frames represent sublattices $S_{\mathcal{D},\mathcal{A},\mathcal{F}}$ consisting of qubits only within a distance of $2$ from the center data, ancilla, or flag qubit, respectively.
    The red labels of the qubits represent the qubit correspondence between (a) and (b).
    (a)
    Geometrical layout of the qubits and simultaneous CR drives on the heavy-hexagonal lattice.
    Arrows represent CR drives from the control to the target qubit, where the label $i$ belongs to the scheduling $P_i$.
    (b)
    Quantum circuit representations of the syndrome extraction procedures on the sublattices $S_{\mathcal{D},\mathcal{A},\mathcal{F}}$.
    Each circuit consists of the simultaneous CR drives $\qty{P_i}_{i=0}^{6}$ separated by black dotted lines and the parity measurements.
    The rectangles labelled with $\mathrm{M}_{Z,X}$ represent the projective measurements on the respective axes.
    If the control or target qubits are outside the sublattices, the other end of the CNOT gate is shown as the open end.
    }
    \label{fig:demo_hhl}
\end{figure*}

In \cref{sec:demo_cr}, we analyzed the CR gate on an isolated system, i.e., a mono-modal drive on a system with a few number of qubits.
In this subsection, we analyze the heavy-hexagon code~\cite{PhysRevX.10.011022, neeraja2022matching}, which is a subsystem stabilizer code~\cite{PhysRevA.73.012340, PhysRevLett.98.220502, PhysRevLett.95.230504}, with qubits forming a heavy-hexagonal lattice as in \cref{fig:demo_hhl}~(a).
We note that the Floquet analysis is substantially more involved due to the multi-modal nature of the problem.

In the heavy-hexagon code, qubits belong to one of the data~($\mathcal{D}$), ancilla~($\mathcal{A}$), or flag~($\mathcal{F}$) qubits shown as yellow, white, or blue circles, respectively.
The data qubits are aligned in a square lattice, as shown in \cref{fig:demo_hhl}~(a).
Let us assume a $d\times d$ square lattice consisting of the data qubits, where $\sigma_{i,j}$ denotes the Pauli operator of the data qubit in the $j$th row and the $i$th column.
The gauge and stabilizer groups of the heavy-hexagon code are given as follows:
\begin{align}
\nonumber \mathcal{G}_{\mathrm{hex}}=& \langle Z_{i,j}Z_{i+1,j}, \ X_{i,j}X_{i,j+1}X_{i+1,j}X_{i+1,j+1}, \\
& \quad  X_{1,2m-1}X_{1,2m},  \ X_{d,2m}X_{d,2m+1} \rangle \;, \\
\nonumber \mathcal{S}_{\mathrm{hex}}=&\langle  Z_{i,j}Z_{i,j+1}Z_{i+1,j}Z_{i+1,j+1}, \
 Z_{2m, d}Z_{2m+1, d}, \\
 &Z_{2m-1, 1}Z_{2m, 1}, \ \prod_i X_{i,j}X_{i,j+1}  \rangle \;,
\end{align}
where $m\in[1, 2, \cdots, \frac{d-1}{2}]$, and $i+j$ is constrained to be even for the second term in the gauge group and odd for the first term in the stabiliser group.
Figure~\ref{fig:demo_hhl}~(a) shows a geometrical layout of the simultaneous CR drives named $\qty{P_i}_{i=0}^{6}$ in an error syndrome extraction procedure of the heavy-hexagon code.
Arrows represent the CR drives in the syndrome extraction, from the control to the target, where those with index $i$ belong to the simultaneous CR drives $P_i$, and are executed simultaneously.

In principle, each qubit in the heavy-hexagon lattice can have different parameters, which makes the collision anlaysis too high a degree of freedom to discuss.
Therefore, we focus on the topological analysis of the Floquet Hamiltonian of the heavy-hexagon code, and discuss potential Floquet~(frequency) collisions in a parameter-independent way.
Here, we count up to the second-order potential Floquet~(frequency) collisions in the heavy-hexagon code by applying \Cref{collision_finder_w_structure}.
A potential $k$th-order Floquet collision involving the center qubit is defined as a pair of Floquet states satisfying the following conditions:
\begin{enumerate}
\item\label{cond1} Pairs that are mapped by translation of the $\mathrm{BZ}$ index are treated as identical~(\cref{sec:floquet}),
\item\label{cond2} The pair must contain at least one computational state~(\cref{sec:complexity}),
\item\label{cond3} The pair must be coupled via a walk of length $k$ in the Floquet Hamiltonian~(\cref{sec:floquet_space}),
\item\label{cond4} The walk must be valid and involves the center qubit~(\cref{sec:real_space}).
\end{enumerate}
The corresponding potential $k$th-order frequency collisions are then defined as the frequency conditions under which bare Floquet quasi-excitation energies of the any two of the Floquet states in the valid walk of the $k$th-order Floquet collision become degenerate.

According to \cref{sec:real_space}, for the quantitative calculation of the values of the collision angles of the Floquet collisions involving a center qubit, we have to consider the sublattice consisting of qubits within distance $\lfloor{3k/2\rfloor}$.
For counting the number of the up to $k$th-order potential Floquet~(frequency) collisions, however, the distance is reduced to $k$, because we don't need to calculate the energy shifts on the colliding Floquet states, but only search for the valid walks of length $k$ between them.
In \cref{fig:demo_hhl}~(a), red frames represent sublattices $S_{\mathcal{D}, \mathcal{A}, \mathcal{F}}$ consisting of qubits only within a distance of $2$ from the center data, ancilla, or flag qubit, respectively.
\Cref{fig:demo_hhl}~(b) shows the quantum circuit representations of the syndrome extraction procedures on the sublattices $S_{\mathcal{D},\mathcal{A},\mathcal{F}}$.
If the control or target qubits are outside the sublattices, the other end of the CNOT gate is shown as the open end.
As shown in \cref{fig:demo_hhl}~(b), the simultaneous CR drives correspond to mono- to quad-modal Hamiltonian on $7$-qubit systems.

Consider the operation basis of simultaneous CR drives.
Suppose $n$ simultaneous CR drives, with the $i$th CR drive between the control qubit $c_i$ and target qubit $t_i$.
In the heavy-hexagon code, the control and target qubits of each CR drive do not overlap with those of other CR drives.
All CR drives have different drive frequencies and are assigned different $\mathrm{BZ}$ index $n_i$.
Therefore, the operation basis of the simultaneous CR drives is given as the tensor product of the operation basis of the $i$th CR drive~(\cref{eq:cr_operation}):
\begin{align}
\ket{\va{c},\va{t};\va{n}}=\bigotimes_{i}\ket{c_i,t_i;n_i} \;.
\end{align}
In the heavy-hexagon code, there are also spectator qubits, which should be sufficiently detuned from any CR drives, and hence idle under ideal control.
The operation basis with spectator qubits $\va{s}$ is written as follows:
\begin{align}
\ket{\va{c},\va{t},\va{s};\va{n}}=\left(\bigotimes_{i}\ket{c_i,t_i;n_i}\right)\otimes\left(\bigotimes_{j}\ket{s_j}\right) \;,
\label{eq:simul_cr_operation_basis}
\end{align}
where $s_j \in [g,e,f,\cdots]$ represents the $j$th spectator qubit state.
Note that even in the case where multiple CR drives share their control or target qubits~\cite{heya2018variational, PRXQuantum.2.040336, kim2022high}, we can define the operation basis as an entangling Floquet state between qubits and analyze Floquet collisions.
However, for systems with non-local operation basis, the sublattice extraction procedure in \Cref{collision_finder_w_structure} requires some adjustments, which is discussed in \Cref{sec:truncated_sys}.

Following the procedures of \Cref{collision_finder_w_structure}, we apply \Cref{collision_finder_wo_structure} to simultaneous CR drives $\qty{P_i}_{i=0}^{6}$ on the sublattices $S_{\mathcal{D}, \mathcal{A}, \mathcal{F}}$ as a subroutine of \Cref{collision_finder_w_structure}.
The numerical calculation consists of the following procedures.
First, an empty graph $G_F$ is created.
Based on the given sublattice $S_{x}$ and the simultaneous CR drives $P_{y}$, $2^7$-computational states are computed as the operation basis, based on \cref{eq:simul_cr_operation_basis}, and added to the graph $G_F$ as nodes.
Note that due to the translational symmetry of the Floquet Hamiltonian with respect to the $\mathrm{BZ}$ index, the same results are obtained for any choice of the $\mathrm{BZ}$ indices of the initial computational states.
Since the operation basis is in a tensor product state, even if one of the control and target qubits is outside the sublattice, the operation basis of the other qubit remains intact.
The exchange interactions and microwave drives in the sublattices are transformed into perturbations in the Floquet Hamiltonian according to \cref{eq:fourier_series_H_expansion}, and regarded as the edges in the graph $G_F$.
We then extract the region which is reachable by tracing the edges $k$ times from the initial nodes.
Finally, in this region, we search for up to second-order potential frequency and Floquet collisions.

\renewcommand{\arraystretch}{1}
\begin{table*}
	\caption{
	Numerical simulations of \Cref{collision_finder_w_structure} for syndrome extraction procedures on the heavy-hexagon code.
    The vertical and horizontal axes of the table correspond to the sublattices $S_{\mathcal{D}, \mathcal{A}, \mathcal{F}}$ and simultaneous CR drives $\qty{P_i}_{i=0}^{6}$ subject to \Cref{collision_finder_w_structure}, respectively.
    In each column, $n_F^{(k)}$ and $n_f^{(k)}$ are the number of $k$th-order Floquet~(frequency) collisions, respectively.
    More detailed discussion is provided in \cref{sec:demo_qec}.
	}
	\label{tab:collision_qec}
	\begin{tabular}{p{1cm}p{1cm}p{1cm}p{1cm}p{1cm}p{1cm}p{1cm}p{1cm}p{1cm}p{1cm}p{1cm}} \hline \hline
	            &               & $P_0$ & $P_1$ & $P_2$ & $P_3$ & $P_4$ & $P_5$ & $P_6$ \\
	    \hline
	    $S_{\mathcal{D}}$
	           & $n_{F}^{(1)}$ & 416   & 416   & 640   & 1344  & 416   & 640   & 640   \\
	           & $n_{F}^{(2)}$ & 2432  & 4224  & 5312  & 9728  & 2432  & 4992  & 4992  \\
	           & $n_{f}^{(1)}$ & 6     & 6     & 6     & 6     & 6     & 6     & 6     \\
	           & $n_{f}^{(2)}$ & 77    & 77    & 71    & 94    & 77    & 97    & 97    \\
	   \hline
	    $S_{\mathcal{A}}$
	           & $n_{F}^{(1)}$ & 416   & 416   & 192   & 416   & 416   & 640   & 640   \\
	           & $n_{F}^{(2)}$ & 2432  & 4224  & 1920  & 4224  & 2432  & 4992  & 4992  \\
	           & $n_{f}^{(1)}$ & 6     & 6     & 6     & 6     & 6     & 6     & 6     \\
	           & $n_{f}^{(2)}$ & 77    & 77    & 57    & 77    & 77    & 97    & 97    \\
	   \hline
	    $S_{\mathcal{F}}$
	           & $n_{F}^{(1)}$ & 288   & 1664  & 736   & 512   & 960   & 960   & 960   \\
	           & $n_{F}^{(2)}$ & 1728  & 10624 & 4736  & 4480  & 4224  & 7488  & 7488  \\
	           & $n_{f}^{(1)}$ & 9     & 9     & 9     & 9     & 9     & 9     & 9     \\
	           & $n_{f}^{(2)}$ & 63    & 110   & 87    & 75    & 95    & 95    & 95    \\
	    \hline \hline
	\end{tabular}
\end{table*}
\Cref{tab:collision_qec} shows the results of the numerical Floquet analysis.
The indices $n_F^{(k)}$ and $n_f^{(k)}$ represent the number of potential $k$th-order Floquet~(frequency) collisions, respectively.
An interesting observation is that two symmetric sequences have the same number of Floquet~(frequency) collisions.
Here, symmetric sequences are those that are equivalent by mirroring while preserving the center qubit.
In particular, $P_0$ and $P_4$, $P_5$ and $P_6$ in $S_{\mathcal{D}}$, and $P_0$ and $P_4$, $P_1$ and $P_3$, $P_5$ and $P_6$ in $S_{\mathcal{A}}$ fall under the symmetric sequences.
Moreover, from \Cref{tab:collision_qec} and \cref{fig:demo_hhl}~(a), we find that the number of first- and second-order Floquet collisions depends only on the position of the target qubits within the distances $1$ and $2$ from the center qubit, respectively.
As shown in the region within the distance of $2$ from the computational Floquet states in \cref{fig:demo_cr}, there are no Floquet states accessible only via the microwave drives.
Since the number of potential Floquet collisions is only determined by the topological structure of the Floquet energy level diagram, it can be seen that only the position of the target qubits matters.
The microwave drives, however, have a quantitative effect on the collision factors, as found in Tables~\ref{tab:collision_2t} and \ref{tab:collision_3t}.
\Cref{tab:collision_qec} shows also that the number of frequency collisions is much smaller than those of Floquet collisions, suggesting that for a given frequency collision, numerous Floquet collisions occur simultaneously, as in \cref{sec:demo_cr}.
\Cref{tab:collision_qec} shows also that the number of first-order frequency collisions, involving a center qubit, is independent of the sequence and is always three times the number of neighbouring qubits,
since the first-order Floquet collision factor is always non-zero, regardless of the CR drive as found in \Cref{tab:collision_2t} and \ref{tab:collision_3t}.
Therefore, the center qubit always has one type 1 and two type 3 collisions with the neighbouring qubits independently.
On the other hand, the number of second-order frequency collisions depends on the CR drives,
since the corresponding Floquet collision factors can be zero depending on the presence or absence of the CR drive based on Tables~\ref{tab:collision_2t} and \ref{tab:collision_3t}.

The number of potential Floquet~(frequency) collisions for the desired control on a given lattice is a useful quantitative indicator of the difficulty of system frequency allocation.
In principle, we can also calculate the values of the collision angles quantitatively for a given set of system parameters.
In reality, system parameters may get finite variation due to fabrication imperfection~\cite{hertzberg2020laserannealing}.
Such parameter variation can be also considered by Monte Carlo sampling~\cite{mackay1998introduction} and our method can provide robust parameter design for the syndrome extraction operation of the heavy-hexagon code.
Our current implementation takes about several hours to check up to second-order Floquet collisions involving a particular qubit in the heavy-hexagon code for a given system parameter.
However, we expect that methods compatible with high performance computing, such as tensor networks~\cite{orus2014practical}, can drastically reduce the computational time in the future.

\section{Summary and Discussion} \label{sec:conclusion}
In this paper, we employ Floquet theory to the analysis of frequency collisions.
We quantitatively formulate Floquet collisions by collision angles between Floquet states in the operation basis.
We show that the collision angles can be calculated with finite computational complexity using a perturbative approach.
In the perturbative analysis of Floquet collisions, we introduce a collision order and show the relation between lower bound on the collision order and the distance in Floquet and real space.
Using this relation, we propose an efficient collision analysis method for general and sparse qubit lattices and estimate their computational complexities.

We apply these methods to relevant experimental situations.
First, for an ideal two-transmon qubit CR gate and a three-transmon extension, we performed analytical perturbative calculations of the collision bounds, which have been investigated numerically so far~\cite{hertzberg2020laserannealing}.
Next, we observed the nature of the Floquet~(frequency) collisions through numerical simulations.
The simulations show two overlooked collision mechanisms: the type 8 collision and the collision avoided crossing.
Type 8 is the collision caused by the $ZZ$ interaction between qubits, which implies that unless the $ZZ$ interaction is eliminated~\cite{kandala2020demonstration, PhysRevLett.129.060501}, CR gates always have the Floquet collision at the pulse edges, regardless of the control-target detuning.
We also show that the rotary tone~\cite{PRXQuantum.1.020318} can modulate the Floquet Hamiltonian and potentially mitigate the type 8 collision.
Collision avoided crossing is the anti-crossing effect between the frequency collisions derived from simultaneous Floquet collisions.
It suggests that frequency allocation in large-scale quantum processors requires additional attention compared to the conventional frequency allocation in small-scale quantum processors~\cite{hertzberg2020laserannealing, doi:10.1126/sciadv.abi6690}.
Finally, we analyzed the more complex problem of frequency collisions in heavy-hexagon codes, where we give a quantitative estimation of the difficulty of the system frequency allocation.

Our analysis imposed several approximations to the considered system and control.
For a more realistic analysis, we have the following prospects.
First, we modeled transmon qubits as weakly nonlinear Duffing oscillators, as the leading approximation of the Josephson nonlinearity~\cite{PhysRevA.76.042319}.
In particular, the higher excited levels of transmon are more sensitive to charge dispersion~\cite{PhysRevA.76.042319, PhysRevB.77.180502}.
Therefore, collisions that involve such high-energy excited states, e.g. due to high-power off-resonant drive in dispersive protocols~\cite{PhysRevApplied.11.024003, PhysRevApplied.11.014030, PhysRevApplied.18.034031, PhysRevA.105.022607, cohen2022reminiscence,https://doi.org/10.48550/arxiv.2212.05097}, are very sensitive to charge noise.
Transmon qubits also have finite coherence times due to coupling to the environment and noise.
Since transitions are generally broadened by the finite coherence time, collision bounds are also expected to be broadened.
Some extensions of the Floquet~\cite{sato2020floquet} and perturbation theory~\cite{PhysRevA.106.052601} are applicable to open systems.
Taking into account such overlooked features will allow for a more precise characterization of frequency collisions.

Second, we approximated control signals as CW drives, while in practice the control signals have a finite duration pulse envelope.
Control signal envelopes lead to a time variation of the effective Hamiltonian~\cite{malekakhlagh2021mitigating}.
The steep time variation of the Hamiltonian induces non-adiabatic transitions between the instantaneous eigenstates~\cite{born1928beweis,kato1950adiabatic}.
A closed loop in the control parameter space is known to induce a Berry phase to the final state~\cite{berry1984quantal}.
Similar phenomena is also reported for the Floquet Hamiltonian~\cite{PhysRevA.56.4045, uchida2022diabatic}.
A Floquet analysis, while incorporating the control signal envelopes, will provide robust control against frequency collisions.

\section*{Acknowledgements}\label{sec:acknowledgements}
We acknowledge Ted Thorbeck for a fruitful discussion on a relation between charge dispersion and frequency collision, Takashi Imamichi for a fruitful discussion on qubit frequency allocation, Toshinari Itoko for a comment on the relation between collision and real space distance, Shuhei Tamate for a fruitful discussion on block diagonalisation, Yutaka Tabuchi for a fruitful discussion on Floquet and perturbation theory, and Yasunobu Nakamura for a fruitful discussion on a research direction.

\section{Author contributions}
K.H. designed the theoretical concepts.
K.H. performed numerical and analytical calculations.
K.H. and S.M. proved theorems in the paper.
M.M. assessed the validity of the proposed method by comparing with previous studies.
K.H. wrote the manuscript with feedback from the other authors.
N.K. and E.P. supervised the project.

\section{Data availability}
The data that support the findings of this study are available from the corresponding authors upon reasonable request.

\section{Code availability}
The code that is deemed central to the conclusions are available from the corresponding author upon reasonable request.

\section{Competing interests}
The authors declare no competing interests.

\appendix

\section{Numerical test of the distance law of the collision orders}
\begin{figure*}
    \centering
	\includegraphics[width=0.9\textwidth]{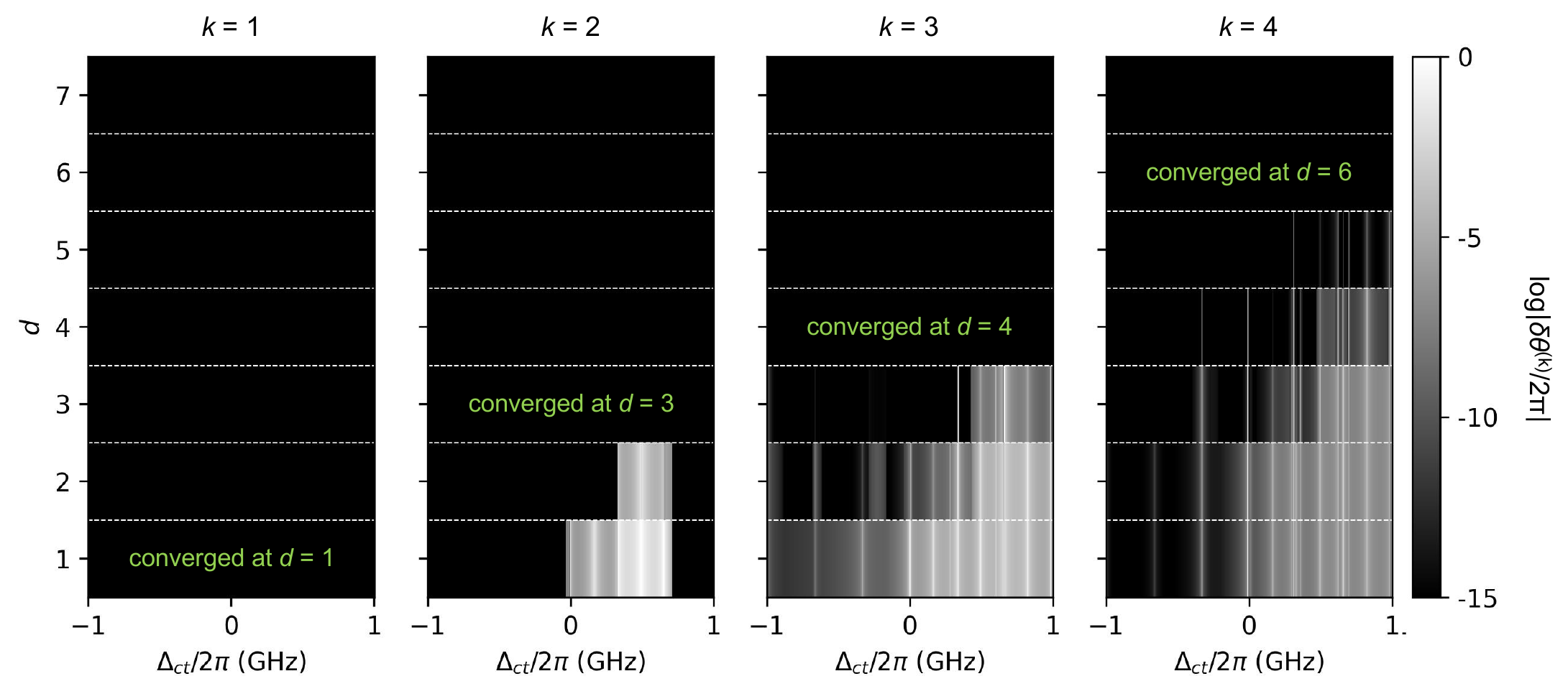}
    \caption{
    Numerical tests of the distance law of the collision order in \cref{sec:floquet_space},
    for a two-transmon system with a CR drive. 
    Each figure shows the gray-scale log plot of the convergence error $\delta \theta^{(k)}$ of the $k$th-order Floquet collisions from $k=1$ to $4$.
    In each figure, we perform a 2D sweep of the control-target detuning $\Delta_{\mathrm{ct}}/2\pi$ from $-1$ to $1$~GHz and the collision search distance $d$ from $1$ to $7$.
    Qubits anharmonicity and qubit-qubit coupling are set to $-330$~MHz and $3.8$~MHz, respectively.
    }
    \label{fig:distance_rule}
\end{figure*}
In \cref{sec:floquet_space}, we proved an analytical distance law of the collision order on the Floquet space.
In this appendix, we re-examine the distance law using numerical simulations.
Let us assume the CR drive on a two-transmon system, as in \cref{sec:demo_cr}.
In \cref{sec:demo_cr}, we analyzed the $k$th-order Floquet collision from the Floquet subspace within the search distance $d=\lfloor{3k/2\rfloor}$ from the computational states.
In this appendix, we numerically check the convergence of the collision angles while sweeping the search distance.
In the simulation, we define the convergence error $\delta \theta^{(k)}$ as the difference between the $k$th-order collision angle derived with a sufficiently large search distance ($d=7$) and with a smaller search distance ($d\leq6$).
\Cref{fig:distance_rule} shows a gray-scale log plot of the convergence error for up to the fourth-order Floquet collisions with 2D sweep of the control-target detuning and the search distance.
\Cref{fig:distance_rule} shows that the convergence error decreases monotonically with increasing the search distance and converges to zero at a particular search distance corresponding to the collision order.
From \cref{fig:distance_rule}, we can find that the search distance $d=\lfloor{3k/2\rfloor}$ is sufficient to converge $\delta \theta^{(k)}=0$ for any control-target detuning $\Delta_{\mathrm{ct}}$ as predicted in \cref{sec:floquet_space}.

\section{Collision analysis for sparse qubit lattice with non-local operation basis} \label{sec:truncated_sys}
\begin{figure*}
\begin{minipage}{\linewidth}
\begin{algorithm}[H]
\caption{Collision analysis for sparse qubit lattice with non-local gates} 
\label{collision_finder_w_nlgates}
\begin{algorithmic}[1]
    \State Regard qubit subsets sharing the non-local operation basis as a single node
    \State Construct a graph $G'_R$ consisting of such nodes
    \For {center node from the graph $G'_R$}
        \State Extract a sublattice including only the nodes within distance $\lfloor{3k/2\rfloor}$ from the center node
        \State Construct a subspace Hamiltonian $H^{\mathrm{sub}}$ corresponding to the sublattice
        \For {qubit from the center node}
            \State Verify $k$th-order Floquet collisions involving the qubit with \Cref{collision_finder_wo_structure}
        \EndFor
        \State Remove the center node from the graph $G'_R$
    \EndFor
\end{algorithmic}
\end{algorithm}
\end{minipage}
\end{figure*}
\begin{figure*}
    \centering
	\includegraphics[width=0.8\textwidth]{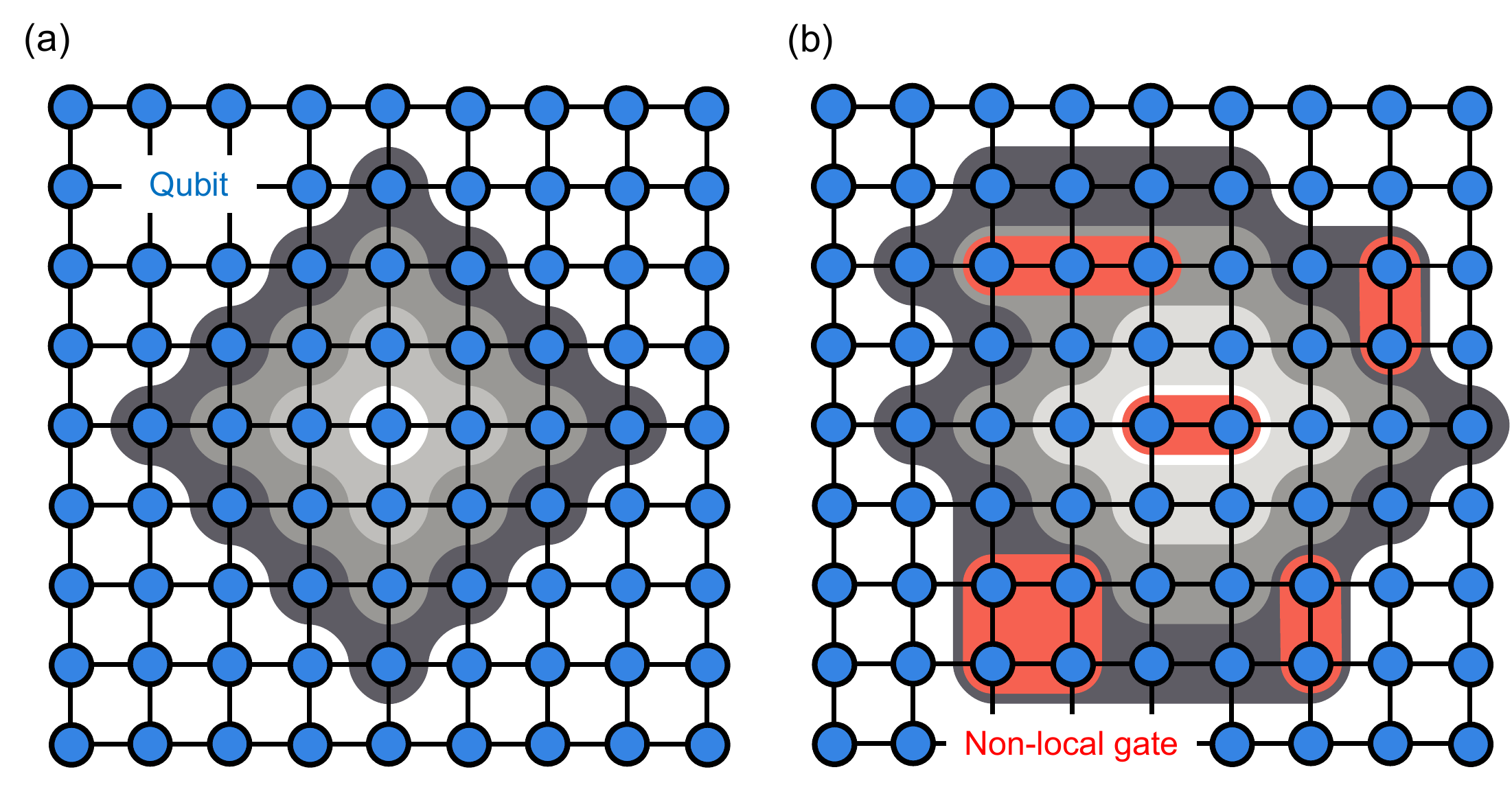}
    \caption{
    Schematic of the distance law of the perturbation order on the sparse qubit lattices without~(a) and with~(b) the non-local gates.
    The blue circles and black lines represent the qubits and the couplings between them, respectively.
    (a) The sparse qubit lattice without the non-local gates.
    The grey regions represent the order of the perturbation derived from the center qubit, with darker shades indicating higher order, which depends only on the qubit lattice structure.
    (b) The sparse qubit lattice with the non-local gates.
    Each subset of qubits grouped in the red region is the target qubit of the each non-local gate.
    The grey regions represent the order of the perturbation derived from the center qubit, with darker shades indicating higher order, which depends both structures of the qubit lattice and non-local gates.
    }
    \label{fig:nlop}
\end{figure*}
In this paper, we have mainly focused on two-qubit gates with a local operation basis, such as the CR gates.
As shown in \cref{eq:cr_operation}, the operation basis of the CR gate can be written by tensor product states for each qubit.
The same applies to othere local gates such as CZ and Toffoli gates.
On the other hand, the operation basis of the non-local gates, such as iSWAP and SWAP gates, form entangling states between qubits.
For example, in the case of the cross-cross resonance gate~\cite{PRXQuantum.2.040336}, the operation basis in the computational subspace is given as
\begin{align}
\ket{\psi_{0}} &= \frac{1}{2}\qty(-\ket{00}+\ket{01}+\ket{10}+\ket{11}) \;, \\
\ket{\psi_{1}} &= \frac{1}{2}\qty(+\ket{00}-\ket{01}+\ket{10}+\ket{11}) \;, \\
\ket{\psi_{2}} &= \frac{1}{2}\qty(+\ket{00}+\ket{01}-\ket{10}+\ket{11}) \;, \\
\ket{\psi_{3}} &= \frac{1}{2}\qty(+\ket{00}+\ket{01}+\ket{10}-\ket{11}) \;,
\end{align}
where we have omitted the $\mathrm{BZ}$ indices for simplicity.
In such cases, we should be careful about the sublattice extraction procedure in \Cref{collision_finder_w_structure}, because the qubits sharing the entangling operation basis are indivisible.
We summarise our Floquet-based collision analysis for the sparse qubit lattice with non-local gates in \Cref{collision_finder_w_nlgates}.
In \Cref{collision_finder_w_nlgates}, we first map a subset of qubits sharing the entangling operation basis as a single collective node.
Then, we reconstruct the graph $G'_R$ from such nodes, where each collective node couples to all the nodes originally coupled to the qubits in that node.
Next, we select the center node of the graph $G'_R$ and extract the sublattice consisting only of nodes within a distance $d=\lfloor{3k/2\rfloor}$.
The rest of the procedure is the same as in \Cref{collision_finder_w_structure}.
\Cref{fig:nlop} shows an overview of the sublattice extraction in \Cref{collision_finder_w_nlgates}.

Consider the computational complexity of \Cref{collision_finder_w_nlgates}.
If the entire system is contained in a single many-body non-local gate, the computational complexity of \Cref{collision_finder_w_nlgates} is equal to \cref{eq:computational_complexity_wo}.
On the other hand, for the case where the system is filled with at most $x$-body non-local gates, the computational complexity $C_3(k,n,d,m,r,x)$ is calculated as follows.
From the discussion in \cref{sec:complexity}, consider a system of $n$-qudits with $d$-levels, $m$-drives and degree $r$.
The graph $G'_R$ corresponding to the system then consists of $n/x$ nodes and has degree $xr$.
Each nodes can be regarded as $d^x$-level system with $xm$-drives.
Therefore, the computational complexity $C_3(k,n,d,m,r,x)$ is given as follows:
\begin{align}
C_3(k,n,d,m,r,x)
&=C_2 \qty(k,\frac{n}{x},d^x,xm,xr)\;.
\label{eq:complexity-3}
\end{align}
\Cref{eq:complexity-3} increases exponentially with the parameter $x$ and suggests that the presence of non-local gates tends to make the collision analysis more difficult.

\bibliography{reference.bib}

\end{document}